\documentclass[12pt]{article}
\usepackage[english]{babel}
\usepackage{ulem}
\usepackage[utf8]{inputenc}
\usepackage[T1]{fontenc}
\usepackage{graphicx}
\usepackage{cancel}
\usepackage{amsmath,amssymb,amsfonts,amsthm}
\usepackage{geometry}
\usepackage{mathrsfs}
\usepackage{nicefrac}
\usepackage{booktabs}
\usepackage{multirow}
\usepackage{enumitem}
\usepackage{float}
\usepackage{hyperref}
\usepackage{blkarray}
\hypersetup{colorlinks = true, linkcolor=red, citecolor=blue}
\usepackage{mathtools}
\usepackage{upgreek}

\usepackage[ruled]{algorithm2e}
\usepackage{authblk}
\usepackage{subcaption}
\usepackage{bm} %bold math font
\usepackage{todonotes}
\DeclareUnicodeCharacter{202F}{FIX ME!!!!}
\newcommand{\R}{\mathbb R}
\newcommand{\N}{\mathbb N}

\newcommand{\C}{\mathbb C}

\newcommand{\cO}{{\mathcal O}}

\newcommand\norm[1]{\lVert #1 \rVert}

\newcommand\inner[1]{\langle #1 \rangle}

\newcommand{\rhobar}{ \overline{\rho}} %   {\rm {det}}}
\DeclareMathOperator{\rank}{rank}
\DeclareMathOperator*{\esssup}{ess\,sup}

\DeclareMathOperator{\ran}{ran}

\newcommand\scpr{\boldsymbol{\cdot}}

\newcommand{\ra}{\rightarrow}

\begin{document}
\numberwithin{equation}{section}
\newtheorem{theo}{Theorem}[section]
\newtheorem{prop}[theo]{Proposition}
\newtheorem{note}[theo]{Remark}
\newtheorem{lem}[theo]{Lemma}
\newtheorem{cor}[theo]{Corollary}
\newtheorem{definition}[theo]{Definition}
\newtheorem{assumption}{Assumption}
%

%\title{\textcolor{red}{Time-dependent density functional theory in the linear response regime: mathematical foundations and pole shifting}} 

\title{The density-density response function in time-dependent density functional theory: mathematical foundations and  pole shifting} 

%\title{Pole-shifting of the linear response function in time-dependent density functional theory}
\author{Thiago Carvalho Corso\thanks{Email: \url{thiago.carvalho@ma.tum.de}}}
\affil{Zentrum Mathematik, Technische Universit\"at M\"unchen, Germany}
\author{Mi-Song Dupuy\thanks{Corresponding author, email: \url{mi-song.dupuy@sorbonne-universite.fr}}}
\affil{Laboratoire Jacques-Louis Lions, Sorbonne Université, Paris, France}
\author{Gero Friesecke\thanks{Email: \url{gf@ma.tum.de}}}
\affil{Zentrum Mathematik, Technische Universit\"at M\"unchen, Germany}
%\thanks{}
%(Email: \url{dupuy@ma.tum.de})}}

%\emails{dupuy@math.univ-paris-diderot.fr}
%\affil{Univ. Paris Diderot, Sorbonne Paris Cit\'e, Laboratoire Jacques-Louis Lions, UMR 7598, UPMC, CNRS, F-75205 Paris, France, {dupuy@math.univ-paris-diderot.fr}}

\renewcommand\Affilfont{\itshape\small}

\maketitle

%\begin{abstract}
%    The goal of this paper is two-fold: in the first part we define and identify some properties of the retarded linear response function for typical non-relativistic electronic Hamiltonians for molecules and atoms. In particular, we characterize their Fourier transform and show that they are integral operators, with explicit estimates on their kernel, away from the countable set of excitation frequencies corresponding to eigenvalues and thresholds. In the second part of the paper we focus on the Dyson equation stemming from the Random Phase Approximation (RPA) from Time-dependent Density Functional Theory (TDDFT). We start by showing existence, uniqueness and invertibility of the Dyson equation for the linear response function in a suitable space of operator-valued functions. Moreover, using the properties derived in the first part, we show that solutions from the Dyson equation whose initial conditions correspond to linear response function of our model Hamiltonians are tempered distribution and that all poles, \textit{before the ionization threshold}, of its Fourier transform are forward shifted with respect to the poles from the initial linear response function.
%\end{abstract}
\begin{abstract}
    We establish existence and uniqueness of the solution to the Dyson equation for the 
    density-density response function in time-dependent density functional theory (TDDFT) in the random phase approximation (RPA). 
    We show that the poles of the RPA density-density
    response function are forward-shifted with respect to those of the non-interacting 
    response function, thereby explaining mathematically the well known empirical fact that the non-interacting poles (given by the spectral gaps of the time-independent Kohn-Sham equations) underestimate the true transition frequencies.
    %the RPA linear response function has a Fourier transform with poles that are forward-shifted compared to the noninteracting linear response function.
    %We prove that the poles of the RPA linear response function 
    Moreover we show that the RPA poles are solutions to an eigenvalue problem, justifying the approach commonly used in the physics community to compute these poles.

\end{abstract}
\tableofcontents

\section{Introduction}

{While ground state properties of molecules are very successfully captured by time-independent Kohn-Sham density functional theory (KS-DFT), 
excitation energies provide a much greater challenge. In particular, the excitation energies of the time-independent Kohn-Sham equations do not accurately capture the true excitation energies, and {\it have no theoretically supported meaning}.}

{Instead, time-dependent density functional theory (TDDFT) in the linear response regime has been found to capture a molecule's excitation spectrum much more accurately (see e.g. \cite{Vasiliev02}). The underlying Dyson equation for the density-density response function of TDDFT has been derived as a meaningful approximation for this task (see e.g. \cite{marques2012fundamentals,ullrich2012time}), and the overall approach has a huge physics literature. Our goal in this paper is to put TDDFT in the linear response regime and its connection with excitation spectra on a firm mathematical footing. We  
\begin{itemize}
\item[(1)] 
establish existence and uniqueness of a solution to the Dyson equation for the density-density response function, in the basic case of the random phase approximation (RPA)
%we are not aware of any previous rigorous results on the Dyson equation, neither in the RPA nor any other approximation
\item[(2)] 
mathematically clarify the relationship between the density-density response function and excitation spectra, by proving that the 'exact' response function (coming from the evolution of the one-body density under full many-body quantum dynamics) has poles precisely at the excitation frequencies of the many-body Hamiltonian 
\item[(3)] show that the excitation frequencies obtained from the RPA Dyson equation are always forward shifted with respect to those of the time-independent Kohn-Sham equations. 
\end{itemize}
}

{Here, (1) proceeds by naturally viewing the density-density response function (DDRF) at a given time $t$ as a linear operator between one-body potentials and identifying a suitable class of potentials on which the Dyson equation is well posed.} 

{(2) is considered 'well known' in the physics literature. But the underlying Lehmann representation of the DDRF is not strictly speaking applicable to the molecular Hamiltonians to which one seeks to apply it in practice, as it tacitly assumes purely discrete spectrum and misses contributions from the continuous spectrum. Our advance is to provide a rigorous Lehmann representation of the DDRF which applies to molecular Hamiltonians and captures the contributions from the continuous spectrum.}

{Finally, (3) proceeds by characterizing  the RPA poles as solutions to a certain eigenvalue problem, and carefully analyzing this eigenvalue problem.}

{Before stating our main results in more detail, let us introduce some background on linear response theory and on TDDFT.}

\subsection{Linear response theory}
Linear response theory allows to compute first-order corrections to quantities of interest of a molecule at equilibrium which is perturbed by an external potential.
The exact wave function $\Psi(t)$ encoding the behaviour of the electrons of the molecules is the solution to the time-dependent Schrödinger equation 
\begin{equation}
    \label{eq:TD Schrodinger}
    \left\lbrace\begin{aligned}
        i \partial_t \Psi(t) &= H(t) \Psi(t), && t>0\\
        \Psi(0) &= \Psi_0 ,
    \end{aligned}\right.
\end{equation}
where 
\begin{equation} \label{ansatz_of_pert}
    H(t) = H + \varepsilon f(t) V_\mathcal{P}
\end{equation}
with $V_\mathcal{P}$ a time-independent bounded multiplicative potential (the \textit{probe potential}) and $f$ a bounded scalar function of time (the \textit{time profile}). $\Psi_0$ is the ground state of the rest Hamiltonian $H$, {which for a molecule has the form}
\begin{align} 
    H = -\tfrac{1}{2}\Delta + \sum_{1 \leq i < j \leq N} w(r_i-r_j) + \sum_{i=1}^N v(r_i), \label{eq:hamiltoniandef0}
\end{align}
where $v,w \in L^2(\R^3)+L^\infty(\R^3)$ and real-valued. %\sout{$H$ is a self-adjoint operator acting on the anti-symmetric subspace $\bigwedge_{i=1}^N L^2(\R^3)$  of $L^2((\R^3)^N)$ with domain $H^2((\R^{3})^N)$.\textcolor{green}{GERO: WHY IGNORE SPIN HERE?} We assume that $H$ has a simple lowest eigenvalue $E_0$ and the corresponding eigenfunction $\Psi_0$ is unique up to a phase factor.}

For an observable $V_\mathcal{O}$ we are interested in the expectation value
\begin{equation}
    \langle V_\mathcal{O} \rangle_t = \langle \Psi(t), V_\mathcal{O} \Psi(t) \rangle.
\end{equation}
Since the perturbation is small, at first order in $\varepsilon$ the variation of $\langle V_\mathcal{O} \rangle_t$ is 
\begin{equation} \label{eq:VOt}
    \langle V_\mathcal{O} \rangle_t = \langle V_\mathcal{O} \rangle_0 + \varepsilon (\mathcal{X}_{V_{\mathcal{O}} V_{\mathcal{P}}} \star f)(t) + \mathcal{O}(\varepsilon^2),
\end{equation}
where $\mathcal{X}_{V_{\mathcal{O}} V_{\mathcal{P}}}$ is the function given by the Kubo formula (see Proposition~\ref{prop:taylor})
\[
    {\mathcal X}_{V_{\mathcal{O}} V_{\mathcal{P}}}(\tau) = -i \theta(\tau) \Big\langle V_\mathcal{O} \Psi_0, e^{-i(H - E_0)\tau} V_\mathcal{P} \Psi_0 \Big\rangle + \mathrm{c.c.}
\]
This function has a Fourier transform, at least in the distributional sense, 
\begin{multline}
  \label{eq:Komega}
  \widehat {\mathcal X}_{V_{\mathcal{O}} V_{\mathcal{P}}}(\omega) = \lim_{\eta \to 0^{+}} \left\langle \Psi_{0}, V_{\mathcal O} \Big(\omega +i\eta - (H-E_0)\Big)^{-1} V_{\mathcal P} \Psi_{0}\right\rangle \\
   - \left\langle \Psi_{0},V_{\mathcal P} \Big(\omega +i\eta + (H-E_0)\Big)^{-1} V_{\mathcal O} \Psi_{0}\right\rangle,
\end{multline}
where $\eta \to 0^{+}$ means the one-sided limit as $\eta$ converges to zero from above.
This formula relates the singularities of the Fourier transform of $\widehat{\mathcal{X}}$ to the spectrum of $H$. More precisely, 
$\widehat{\mathcal X}$ has a pole if $|\omega|$ is an eigenvalue of $H - E_0$. 
When $|\omega|$ belongs to the essential spectrum of $H-E_0$, $\widehat{\mathcal{X}}$ is regular for a wide class of potentials $v$ and $w$ as a result of the limiting absorption principle. 
We refer to \cite{agmon1975spectral,Cyconbook, commutatormethodsbook} for more information on this topic.

The location of the poles of the Fourier transform $\widehat{\mathcal{X}}$ provides access to the spectrum of $H$ and in particular to its low-lying eigenvalues. At first sight, evaluating $\widehat{\mathcal{X}}$ from Equation~\eqref{eq:Komega} is by no mean simpler than diagonalising the many-body operator $H$. However a major simplification can be achieved, at least formally, by (``exact'' and approximate) time-dependent density functional theory (TDDFT), which are time-dependent versions of static Hohenberg-Kohn density functional theory~\cite{Hohenberg1964,lieb1983} respectively static Kohn-Sham density functional theory~\cite{KohnSham1965}. 
Provided the perturbing potential $V_\mathcal{P}$ and the observable $V_\mathcal{O}$ are \textit{one-body potentials} 
\begin{equation} \label{eq:restr}
   V_\mathcal{P} =  \sum_{k=1}^N v_\mathcal{P}(r_k), \quad V_\mathcal{O} =  \sum_{k=1}^N v_\mathcal{O}(r_k),
\end{equation}
the expression for $\mathcal{X}$ becomes
\begin{equation}
    \label{eq:K_to_chi}
    \mathcal{X}_{V_{\mathcal{O}} V_{\mathcal{P}}}(\tau) = \langle v_\mathcal{O}, \chi(\tau) v_\mathcal{P} \rangle,    
\end{equation}
for some universal operator-valued function $\chi$ which is independent of $v_{\mathcal O}$ and $v_{\mathcal P}$ and only depends on the static Hamiltonian $H$ in eq.~\eqref{eq:hamiltoniandef0}. This function, called {\it density-density response function} of $H$, is rigorously constructed in section~\ref{subsec:kubo}. In the physics literature $\chi$ is usually postulated to have an integral kernel, {obtained formally by taking  $v_{\mathcal O}$ and $v_{\mathcal P}$ to be delta functions located at $r$ respectively $r'$,} 
$$
   \chi(r,r',\tau) = \bigl\langle \delta_r,\,\chi(\tau)\delta_{r'}\bigr\rangle,
$$
{and this kernel is known as the density-density response function. The operators $V_{\mathcal P}$ and $V_{\mathcal O}$ in \eqref{eq:restr} are then the density operators $\widehat{\rho}_r=\sum_{i=1}^N \delta_{r}(r_i)$ and $\widehat{\rho}_{r'}=\sum_{i=1}^N \delta_{r'}(r_i)$, explaining the name.} Mathematically, there is no need for -- or advantage from -- such an integral representation; $\chi$ is simply an \textit{operator-valued function of time acting on one-body potentials}.

The key observation which allows to bring to bear TDDFT is now the following: with the restriction \eqref{eq:restr} to one-body perturbing potentials and one-body observables, the expectation value $\langle V_\mathcal{O} \rangle_t$ only depends on the electronic density $\rho^{\Psi}$ at time $t$,
\begin{equation} \label{rhodef}
    \rho^{\Psi}(t,r) = N\int_{\R^{3(N-1)}}|\Psi(t,r,r_2,\dots,r_N)|^2 \mathrm{d}r_2\dots\mathrm{d} r_N.
\end{equation}
In the following we write $\rho^\Psi(t)$ for the function $\rho^{\Psi}(t,\cdot) \, : \, \R^3\to\R$.
We have 
\begin{equation}
    \label{eq:linear_response_in_rho}
    \langle V_\mathcal{O} \rangle_t = \langle v_\mathcal{O}, \rho^{\Psi}(t) \rangle = \langle v_\mathcal{O}, \rho^{\Psi}(0) \rangle + \varepsilon \langle v_\mathcal{O}, (\chi v_\mathcal{P} \star f)(t) \rangle + \mathcal{O}(\varepsilon^2),
\end{equation}
so by identification, $\chi$ gives the variation of the electronic density to the first order in $\varepsilon$ 
\begin{equation}
    \label{eq:variation_rho}
    \rho^\Psi(t) = \rho^\Psi(0) + \varepsilon (\chi v_\mathcal{P} \star f)(t) + \cO(\varepsilon^2)
\end{equation}
where throughout this paper $a\star f$ denotes convolution in time,  
\begin{equation}\label{convdef}(a\star f)(t)=\int_0^t a(t-s)f(s)\, ds.
\end{equation}
If it is possible to efficiently approximate the density evolution, and hence the density-density response function $\chi$, then we have a way to obtain the variation of the expectation value $\langle V_\mathcal{O} \rangle_t$ in the linear response regime. Such approximations are provided by time-dependent density functional theory (TDDFT). In the next section we only introduce the simplest -- and commonly used --  such approximation, referring to Appendix A for more details and references.

\subsection{Time-dependent density functional theory}

TDDFT aims to reproduce or approximate the evolution of the electronic density $\rho^\Psi$, which is governed by the many-body Hamiltonian $H$, by the evolution of the density of a non-interacting system. More precisely, the electronic density $\rho^\Psi$ is approximated by the electronic density $\rho^\Phi$, where $\Phi$ is the solution to 
\begin{equation}
    \label{eq:TDHartree}
    \left\lbrace
    \begin{aligned}
        i \partial_t \Phi(t) &= H_{\rm eff}(t) \Phi(t), \quad t>0 \\
        \Phi(0) &= \Phi_0
    \end{aligned}\right.,
\end{equation}
for some suitable effective non-interacting Hamiltonian $H_{\rm eff}$.
The initial condition $\Phi_0$ is a Slater determinant whose density $\rho^{\Phi_0}$ approximates the electronic density $\rho^\Psi(0)$ of the exact ground state. In practice it is taken to be the Kohn-Sham determinant, i.e. the ground state of the Hamiltonian \eqref{nonintKSham}.
In this paper {we focus on the physical case of the Coulomb interaction in \eqref{eq:hamiltoniandef0}, $w=\tfrac{1}{|\,\cdot\, |}$,} and the \textit{random phase approximation (RPA),} which corresponds to the effective Hamiltonian
\begin{equation} \label{eq:RPAham}
    H_{\rm eff}(t) = - \tfrac{1}{2}\Delta + \sum_{i=1}^N v(r_i) + v_\mathrm{xc}^{\rm static}[\rho^\Phi(0)](r_i)+ \rho^\Phi(t) * \frac{1}{|\cdot|}(r_i) + \varepsilon f(t) v_\mathcal{P}(r_i). 
\end{equation}
Here $\Phi$ is the solution to \eqref{eq:TDHartree}, $\rho^\Phi$ is its density (eq.~\eqref{rhodef} with $\Psi$ replaced by $\Phi$), $v_\mathrm{xc}[\rho^\Phi(0)]$ is the static Kohn-Sham exchange-correlation potential of the initial density, and $\frac{1}{|\bm{\cdot}|} * \bm{\cdot}$ is the Hartree operator, i.e. the  convolution with the Coulomb potential. Thus in the RPA, the Hartree potential $\rho^\Phi(t)*\frac{1}{|\cdot |}$ (being the dominant part of the interaction) is dynamically updated whereas the exchange-correlation potential is frozen at the initial density. Also updating it dynamically would correspond to the adiabatic local density approximation (ALDA) provided $v_\mathrm{xc}[\rho]$ is given by the LDA.    
%The evolution is nonlinear, but as the initial condition and the operator are separable, it can be reduced to a system of nonlinear equations.
Since we are interested in the linear response regime, it is not necessary to solve the nonlinear system \eqref{eq:TDHartree},\eqref{eq:RPAham}. 
Instead, assuming that $\rho^\Phi$ has a Taylor expansion to order $1$ in $\varepsilon$, it can be shown (see Appendix A) that the variation of $\langle V_\mathcal{O} \rangle_t$ is given by the formula~\eqref{eq:linear_response_in_rho} with $\chi$ replaced by the solution $\chi^\mathrm{RPA}$ to the following Dyson equation
\begin{equation}
    \label{eq:dyson_rpa}
    \chi^\mathrm{RPA}(t) = \chi_0(t) + \Big(\chi_0 \star \big( \frac{1}{|\cdot|} * \chi^\mathrm{RPA}\big)\Big)(t).
\end{equation}
The operator-valued function $\chi_0$ is the density-density response function of the frozen (or static)  Hamiltonian
\begin{align} \label{nonintKSham}
    H_0 = -\frac12 \Delta + \sum_{i=1}^N v(r_i) + v_\mathrm{xc}^{\rm static}[\rho^\Phi(0)](r_i) + \rho^{\Phi}(0)\ast \frac{1}{|\cdot|}(r_i)  . 
\end{align}

Equation~\eqref{eq:dyson_rpa} is called the Dyson equation in the random phase approximation (RPA), and its solution is the RPA density-density response function \cite{marques2012fundamentals,ullrich2012time}. 
This is the simplest approximation of the density-density response function $\chi$ in TDDFT. More sophisticated approximations (see e.g. \cite{maitra2016perspective}) are beyond the scope of the present paper. The interested reader may consult Appendix~\ref{sec:tddft_short_intro} for a compact introduction to TDDFT and the derivation of \eqref{eq:dyson_rpa}, and the monographs~\cite{marques2012fundamentals,ullrich2012time} for an overview of the field from a physics perspective. 

{Finally, we remark that interesting linear response problems arise in many different contexts. For mathematical results on the potential-to-density response of nonlinear Hartree dynamics for crystals see \cite{cances_rpa}. A  mathematical study of the linear response of density and current density in TDDFT to applied electromagnetic fields in a continuum limit can be found in \cite{jianfeng_tddft}.}

\subsection{Main results}

The main focus of the paper is to establish fundamental properties of the solution $\chi^\mathrm{RPA}$ to the Dyson equation~\eqref{eq:dyson_rpa} (including but not limited to existence and uniqueness), under natural assumptions on the Hamiltonian $H$ {associated to the reference density-density} response function $\chi_0$. 

\subsubsection{Assumptions}

We recall here that the rest Hamiltonian considered is
\begin{align} 
    H = -\tfrac{1}{2}\Delta + \sum_{1 \leq i < j \leq N} w(r_i-r_j) + \sum_{i=1}^N v(r_i) \label{eq:hamiltoniandef}
\end{align}
acting on the anti-symmetric $L^2$-space
\begin{align}
    \mathcal{H}_N = \bigwedge_{i=1}^N L^2(\R^3),
\end{align}
where $v,w \in L^2(\R^3) + L^\infty(\R^3)$ are real-valued. Under this condition on $v$ and $w$, it is well-known that the Hamiltonian $H$ is a self-adjoint operator with domain $H^2(\R^{3N})\cap \mathcal{H}_N$ which is bounded from below \cite{ReedSimonI}.

Throughout this paper, we shall also assume that $H$ satisfies the following general assumption.
\begin{assumption} \label{assump:groundstate} Let $v,w \in L^2(\R^3) + L^\infty(\R^3)$ be real-valued functions and $H$ be defined as in \eqref{eq:hamiltoniandef}. Then we assume that
\begin{enumerate}[label = (\roman*)] 
\item\label{it:groundstate} the ground state energy ${E}_0$ of $H$ is a simple isolated eigenvalue;
%\item the ground-state wave function $\Psi_0$ can be chosen real-valued;
\item\label{it:densitybound} the electronic density $\rho^{\Psi_0}$ of the ground state $\Psi_0$ is bounded;
%\item $H$ has eigenvalues of finite multiplicities below its essential spectrum $\sigma_\mathrm{ess}(H)$.
\end{enumerate}
\end{assumption}

Note that assumption~\ref{it:densitybound} is in fact a consequence of \ref{it:groundstate} for Schr\"odinger operators of the form \eqref{eq:hamiltoniandef} (see for instance \cite{simon86schrodinger}). The reason we promote it to an independent assumption here is that all the results from this paper ultimately rely on these two assumptions and not on the specific form of the rest Hamiltonian $H$.

%Assumption $(i)$ is more subtle and, to the knowledge of the authors, no necessary and sufficient conditions on the potentials $v,w$ and the number of particles $N$ are known for it to hold. Alone the existence of a ground state is subtle question and related to the famous Ionization conjecture.  Nonetheless,  it is a typical assumption on linear response theory and we shall not discuss it any further. 
%Although necessary and sufficient conditions for the first assumption to hold are not fully understood, the second  assumption %and third assumptions
%is true for binding molecules, %\emph{i.e.} when $v(r)=-\sum_{k=1}^{N_\mathrm{at}} \frac{Z_k}{|r-R_k|}$ with $Z_k>0$ and $\sum_{k=1}^{N_\mathrm{at}} Z_k \leq N-1$ and $w(r)=\frac{1}{|r|}$ \cite{zhislin1960discussion,zhislin1965spectrum}.

Under the above assumptions we note that the ionization threshold $\Omega$, defined as
\begin{equation}
    \label{eq:ionization_threshold}
    \Omega = \inf \sigma_\mathrm{ess}(H) - E_0,
\end{equation}
is positive. This is a simple but important observation, since most of the results discussed next concern the behaviour of $\widehat{\chi^{\mathrm{RPA}}}$ inside the interval $(-\Omega,\Omega)$.

\subsubsection{Solution to the Dyson equation}

We first show that the Dyson equation~\eqref{eq:dyson_rpa} has a unique solution in 
the space of strongly continuous maps from $\R_+$ to $\mathcal{B}(L^2(\mathbb{R}^3)+L^\infty(\mathbb{R}^3),L^1(\mathbb{R}^3)\cap L^2(\mathbb{R}^3))$, denoted here by
\begin{align}
    C_s\bigr(\R_+;\mathcal{B}(L^2+L^\infty,L^1\cap L^2)\bigr). 
\end{align}
This is the space of density-density response functions of Hamiltonians~\eqref{eq:hamiltoniandef0} as seen in Proposition~\ref{prop:Lpregularity}.
The proof is an application of the Banach fixed point theorem in an appropriate space.

\begin{theo}[Existence of the solution $\chi^{\mathrm{RPA}}$]
  \label{thm:time_rpa_solution}
    Let $\chi_0 \in C_s\bigr(\R_+;\mathcal{B}(L^2+L^\infty,L^1\cap L^2)\bigr)$. Then the following assertions are true:
    \begin{enumerate}[label=(\roman*)]
        \item \label{itthm:existence_rpa} there is a unique solution $\chi^\mathrm{RPA} \in C_s\bigr(\R_+;\mathcal{B}(L^2+L^\infty,L^1\cap L^2)\bigr)$ to the RPA Dyson equation~\eqref{eq:dyson_rpa};
        \item \label{itthm:bijection_rpa} the solution map
            \[
                \mathcal{S}^{RPA} : \left\lbrace
                \begin{aligned}
                    C_s\bigr(\R_+;\mathcal{B}(L^2+L^\infty,L^1\cap L^2)\bigr) &\to C_s\bigr(\R_+;\mathcal{B}(L^2+L^\infty,L^1\cap L^2)\bigr) \\
                    \chi_0 &\mapsto \chi^\mathrm{RPA}
                \end{aligned}\right.
            \]
            is a bijection.
%        \item the Fourier transform of $\chi^\mathrm{RPA}$ is well-defined if and only if $I - F_H^{1/2} \chi_0 F_H^{1/2}$ is invertible, where $F_H^{1/2}$ is the square root of the Hartree operator.
    \end{enumerate}
\end{theo}

\subsubsection{Poles of the RPA density-density response function}

From {eq.}~\eqref{eq:Komega} and {eq.}~\eqref{eq:K_to_chi}, it can be shown (see section \ref{sec:Fourier}) that the Fourier transform of {the density-density response function} $\chi$, defined on $\{ {\rm Im}(z)>0\}$ as $\int_0^\infty \chi(t)e^{izt} \mathrm{d}t$, is an analytic family of operators whose meromorphic extension has simple real poles. This means that for $\omega \in \R$ a pole of $\widehat{\chi}$, there is a neighborhood of $\omega$, an analytic family of operators defined in this neighborhood $z \mapsto K_0(z)$, and a finite-rank operator $K_{-1}$, such that
\begin{equation}
    \widehat{\chi}(z) = K_0(z) + \frac{K_{-1}}{z-\omega}.
\end{equation}
We denote the set of poles of $\widehat{\chi}$ by $\mathcal{P}(\widehat{\chi})$ {and} define the rank of {a}  pole $\omega$ 
%\sout{of $\widehat{\chi}$ denoted by $\text{rank}_\omega(\widehat{\chi})$} 
as \begin{equation}
    \text{rank}_{\omega}(\widehat{\chi}) = \rank K_{-1}.
\end{equation}
{The main result of this paper shows that the RPA density-density response function has a similar structure. More precisely, we show that the Fourier transform of $\chi^{\rm RPA}$ also admits a meromorphic extension whose poles are located along the real axis. 
We also prove that these poles are forward-shifted compared to the poles of the Fourier transform of the reference density-density response function $\chi_0$. }
%\sout{Furthermore, we study the poles of the Fourier transform in detail for the RPA response function $\chi^\mathrm{RPA}$ and prove that they are forward-shifted compared to those of the Fourier transform of $\chi_0$.} 

Note that even the existence of the Fourier transform in the upper half plane is not clear a priori. Indeed, if we 
%\sout{replace the Hartree operator by the identity $I$, i.e. $\chi_0(t) = I$ for any $t\geq 0$} 
{take $\chi_0(t) = K$ for any $t\geq 0$ and some bounded operator $K \in \mathcal{B}\bigr(L^2(\R^3)+L^\infty(\R^3);L^1(\R^3)\cap L^2(\R^3)\bigr)$}, then $\chi_0$ is a strongly continuous map from $\R_+$ to $\mathcal{B}(L^2+L^\infty,L^1\cap L^2)$ but the solution to the Dyson equation~\eqref{eq:dyson_rpa} in this case is given by {$e^{t K F_H} K$}, whose Fourier transform diverges in $\{{\rm Im}(z)< \sup \sigma(K F_H) \}$.
%\textcolor{red}{Thiago: maybe we can leave this last paragraph here out.}

%\textcolor{blue}{
%    Finally we study the poles of $\widehat{\chi^\mathrm{RPA}}$ and in particular how it behaves compared to the poles of the noninteracting linear response function $\widehat{\chi_0}$.
%    We prove that they are forward shifted compared to those of $\chi_0$. Namely if we denote by $0< \omega_1 \leq \omega_2 \leq \dots$ the poles of $\chi_0$ and $0< \omega_1^\mathrm{RPA} \leq \omega_2^\mathrm{RPA} \leq \dots$ the poles of $\chi^\mathrm{RPA}$, then $\omega_k < \omega_k^\mathrm{RPA}$.
%    \todo[inline]{Finir avec l’équation résolue par les poles de chi RPA}
%}

\begin{theo}[Poles of $\widehat{\chi^{\textnormal{RPA}}}$]\label{theo:polesRPA} Let $\chi_0$ be the density-density response function (see the definition in Proposition~\ref{prop:chi_definition}) of a Hamiltonian $H$ satisfying Assumption~\ref{assump:groundstate}. Let $\chi^\mathrm{RPA}$ be the solution of the RPA Dyson equation~\eqref{eq:dyson_rpa} and $\Omega = \inf \sigma_{\textnormal{ess}}(H) - E_0$ be the ionization threshold of $H$. Let $\mathcal{D}_\Omega$ be the set
\begin{align} \label{domainofFT}
    \mathcal{D}_\Omega = \C \setminus \big( (-\infty,-\Omega] \cup [\Omega,\infty) \big);
\end{align}
Then the following holds:
\begin{enumerate}[label =(\roman*)]
\item \label{it:meromorphicpolesRPA} (Meromorphic) The Fourier transform $\widehat{\chi^\mathrm{RPA}} : \mathcal{D}_\Omega \rightarrow \mathcal{B}\bigr(L^2(\R^3)+L^\infty(\R^3),L^1(\R^3)\cap L^2(\R^3)\bigr)$ is a meromorphic family of operators with simple poles contained in $(-\Omega,\Omega)$;
%(see Definition \ref{def:meromorphicfamily});
%\item \label{it:simplerealpolespolesRPA}(Real simple poles) The set of poles of $\widehat{\chi^\mathrm{RPA}}$ in $\mathcal{D}_\Omega$, denoted here by $\mathcal{P}(\chi^\mathrm{RPA})$, is contained in the interval $(-\Omega,\Omega)$. Moreover, all poles are simple and have finite rank.
\item \label{it:forwardshiftpolesRPA}(Forward shift of poles) Let $\mathcal{P}(\widehat{\chi_0})$ be the set of poles of $\widehat{\chi_0}$, then the poles of $\widehat{\chi^\mathrm{RPA}}$ are forward shifted with respect to the poles of $\widehat{\chi_0}$ in the sense that, for any $0<\omega < \Omega$,
\begin{align*}
    \sum_{\substack{\widetilde{\omega} \in \mathcal{P}(\widehat{\chi^\mathrm{RPA}}) \\ |\widetilde{\omega}|<\omega}} \rank_{\widetilde{\omega}}(\widehat{\chi^\mathrm{RPA}}) \leq \sum_{\substack{\widetilde{\omega} \in \mathcal{P}(\widehat{\chi_0}) \\ |\widetilde{\omega}|<\omega}} \rank_{\widetilde{\omega}}(\widehat{\chi_0}).
\end{align*}
%where $\rank_{\omega}(\cdot )$ denotes the rank of the pole $\omega$.
\end{enumerate}
\end{theo}
\begin{note} While the theorem is insensitive to the precise form of the Hamiltonian $H$ governing the reference density-density response function $\chi_0$, the standard choice given in eq.~\eqref{nonintKSham} is of particular physical interest. In this case, the poles are located precisely at the spectral gaps $\epsilon_a-\epsilon_j$ between occupied and unoccupied eigenvalues of the (one-body) Kohn-Sham Hamiltonian 
\begin{equation}
  h_{\rm KS} = -\frac{1}{2} \Delta + v + v_\mathrm{xc}[\rho^\Phi(0)] + \rho^{\Phi}(0)\ast \frac{1}{|\cdot|}
\end{equation} \label{hKS}
of time-independent density functional theory (see Appendix~\ref{appendix:poles_noninteractingDDRF}). Thus statement~\ref{it:forwardshiftpolesRPA} implies that the Kohn-Sham spectral gaps (accounting for multiplicities) always underestimate the excitation frequencies predicted by the RPA. This can be taken as a mathematical explanation of the empirical fact that the Kohn-Sham gaps are commonly too low compared to the experimental excitation frequencies (see e.g. \cite{Vasiliev02}). 

\end{note}
As a last result, we give a rigorous way to find the poles of $\widehat{\chi^\mathrm{RPA}}$, and compute its rank, by solving an eigenvalue problem. This rigorously justifies the standard approach in the quantum chemistry community to find the poles of $\widehat{\chi^\mathrm{RPA}}$, as long as they lie outside the set of poles of $\widehat{\chi}_0$. For the mutual poles of $\widehat{\chi}_0$ and $\widehat{\chi^\mathrm{RPA}}$, the rank can also be computed by solving a similar eigenvalue problem in a reduced space. More precisely, we have the following criteria.
\begin{theo}[Characterization of the poles of $\widehat{\chi^\mathrm{RPA}}$]\label{theo:rankpolesRPA} Let $|\omega| < \Omega$ be a pole of $\widehat{\chi^\mathrm{RPA}}$, {and let $F_H$ be the Hartree operator $f\mapsto \tfrac{1}{|\,\cdot\,|}*f$}. Then the following holds:
\begin{enumerate}[label=(\roman*)]
\item \label{it:polesawayfrompoles} If $\omega$ is not a pole of $\widehat{\chi_0}$, then the rank of $\omega$ as a pole of $\widehat{\chi^\mathrm{RPA}}$ is given by the number of linearly independent solutions in $L^2(\R^3)$ of the equation
\begin{align}
    F_H^{\frac12}\widehat{\chi_0}(\omega)F_H^{\frac12} f = f. \label{eq:eigenvalueeq}
\end{align}
\item \label{it:polesatpoles} If $\omega$ is a pole of $\widehat{\chi_0}$, then the rank of $\omega$ as a pole of $\widehat{\chi^\mathrm{RPA}}$ is given by the number of linearly independent solutions in $L^2(\R^3)$ of
\begin{align}
    P_{V}^\perp F_H^{\frac12}\widehat{\chi_0}(\omega)F_H^{\frac12} P_{V}^\perp f = f, \label{eq:eigenvalueeqatpole}
\end{align}
where $F_H^{\frac12}$ is the square root of $F_H$ given up to a normalization constant by convolution against $\frac{1}{|\cdot|^2}$ and $P_{V}^\perp$ is the orthogonal projection on the orthogonal complement of $V = F_H^{\frac12}\bigr(\ker(H - E_0 + |\omega|)\bigr)$.
\end{enumerate}
\end{theo}

\begin{note} Note that for any solution $f$ of \eqref{eq:eigenvalueeq}, the non-zero function $g = \widehat{\chi_0}(\omega) F_H^{\frac12} f \in L^1(\R^3) \cap L^2(\R^3)$ satisfies the equation $\widehat{\chi}_0(\omega) F_H g = g$. Likewise, a solution $g$ of the latter yields a solution of the former by taking $f = F_H^{\frac12} g \in L^2(\R^3)$. In particular, solving \eqref{eq:eigenvalueeq} is equivalent to solving $g = \widehat{\chi}_0(\omega) F_H g$, which is the starting point of the Casida formalism in TDDFT \cite[Section 3.8]{lin2019mathematical}.
\end{note}

\paragraph*{Strategy of the proof.} The proofs of Theorems~\ref{theo:polesRPA} and \ref{theo:rankpolesRPA} consist of two main steps. The first and most involved step is a detailed spectral analysis of the symmetrized operator $\widehat{\chi}_s =  F_H^{\frac12} \widehat{\chi} F_H^{\frac12}$. This analysis leads to the characterization of the poles of $(1-\widehat{\chi}_s)^{-1}$ in Propositions \ref{prop:1-chi_s_meromorphic} and \ref{prop:forwardshiftchis} below. With these propositions, the second step in our proof is to show that $\widehat{\chi^\mathrm{RPA}}$ is a meromorphic family of operators with simple real poles, and then prove the identity $\rank_\omega (\widehat{\chi^\mathrm{RPA}}) = \rank_\omega \bigr((1-\widehat{\chi}_s)^{-1}\bigr)$ for any $\omega \in (-\Omega,\Omega)$. The key properties of $F_H$ used in our proofs are its positivity and its $L^p$ mapping properties, given by the Hardy-Littlewood-Sobolev inequality. 
In particular, we expect that the results here can be extended to other types of adiabatic approximations used in TDDFT.

\paragraph*{Structure of the paper.} We start by introducing some notation in the next paragraph.
In Section~\ref{sec:gs_linear_response_operator}, we introduce the (exact) density-density response function $\chi$ of a general Hamiltonian and relate it to the celebrated Kubo formula from linear response theory. 
We also derive some $L^p$ smoothing properties of $\chi$ (Lemma \ref{prop:Lpregularity}) and give the formula of its Fourier transform, which can be viewed as a meromorphic family of operators.
In Section~\ref{sec:rpa_in_time}, we prove the existence and uniqueness of solutions to the RPA-Dyson equation~\eqref{eq:dyson_rpa} in the setting of Theorem~\ref{thm:time_rpa_solution}. 
In Section~\ref{sec:symmetrized_chi}, we study the symmetrized operator $\widehat{\chi}_s  = F_H^{\frac12} \widehat{\chi}_0 F_H^{\frac12}$ and characterize the poles of $(1-\widehat{\chi}_s)^{-1}$ in Propositions~\ref{prop:1-chi_s_meromorphic} and \ref{prop:forwardshiftchis}. We then use these propositions to prove Theorems~\ref{theo:polesRPA} and \ref{theo:rankpolesRPA} in Section~\ref{sec:fourier_chi_rpa}.

\subsection*{Notation} 

The set $\R_+ = [0,\infty)$ denotes the set of non-negative real numbers.
For $A$ and $B$ nonnegative scalar quantities, $A \lesssim B$ means that there is an irrelevant positive constant $C$ such that $A \leq C B$. 
We use the following convention for the Fourier transform of functions $f: \R \rightarrow F$ where $F$ is a Banach space
     \begin{align} 
        \widehat{f}(\omega) = \int_\R f(t) e^{i t\omega} \mathrm{d} t \label{eq:fourierconvention1d}.
     \end{align}
Let $F,G$ be Banach spaces, we will denote their respective norms by $\norm{\cdot}_F$ and  $\norm{\cdot}_G$. Moreover, we denote the set of linear continuous operators from $F$ to $G$ by $\mathcal{B}(F,G)$. If $F =G$, we simply use $\mathcal{B}(F)$. The operator norm on $\mathcal{B}(F,G)$ is denoted by
    \begin{align*} 
        \norm{T}_{F,G} = \sup_{\substack{g \in G \\ \norm{g}_{G}=1}} \norm{Tg}_{F}.
    \end{align*}
Whenever it is clear from the context to which operator space $T$ belongs, we shall use only $\norm{T}$ for the operator norm. For an operator $T : F \rightarrow G$ on Banach spaces $F$,$G$, we denote its kernel and range by $\ker T \subset F$ and $\ran T \subset G$. We also use $\rank T = \dim \ran T$ for the rank of $T$.
For $1\leq p \leq \infty$, $L^p(\R^3)$ (or just $L^p$) denotes the standard $L^p$ spaces with respect to Lebesgue measure. We use the notation 
\begin{align*}
    \inner{f, g} = \int_{\R^n} \overline{f(r)} g(r) \mathrm{d}r
\end{align*}
for the standard inner-product on $L^2(\R^n)$.
     We also use $L^p(\R^n) + L^q(\R^n)$ and $L^p(\R^n) \cap L^q(\R^n)$ for the Banach spaces of measurable  functions with finite norms
     \begin{align*} &\norm{f}_{L^p+L^q} = \inf_{f = f_p+f_q} \{ \norm{f_p}_{L^p} + \norm{f_q}_{L^q} \}\\
         &\norm{f}_{L^p\cap L^q} = \max\{ \norm{f}_{L^p},\norm{f}_{L^q}\}.
     \end{align*}
Note if $p$ and $q$ are conjugate exponents, that is to say $p^{-1}+q^{-1}=1$, $L^p+L^q$ is the dual of $L^p\cap L^q$. For a projection $P \in \mathcal{B}(F)$ on a Banach space $F$, i.e. $P^2 = P$, we say that an operator $B \in \mathcal{B}(F)$ is invertible with respect to $P$ if $P BP = B$ and there exists an operator $B^{-1} \in \mathcal{B}(F)$ such that 
\begin{align}
    PB^{-1} P = B^{-1}\quad\mbox{and}\quad B^{-1} B = B B^{-1} = P .
\end{align}
Note that the inverse $B^{-1}$ is unique.

\section{The ground-state density-density response function}
\label{sec:gs_linear_response_operator}

In this section we recall the basics of linear response theory and give a derivation of the density-density response function. We then highlight a few properties of this operator-valued function and give a representation of its Fourier transform.

\subsection{Derivation and Kubo formula}
\label{subsec:kubo}

Let us start with the definition of the linear response function and its connection to the first order variation in the dynamics with respect to some perturbation.

Let $H$ be the static Hamiltonian defined in \eqref{eq:hamiltoniandef} and consider the time-dependent family of self-adjoint operators $H(t)$
\begin{equation}
    \label{eq:TD Hamiltonian}
    H(t) = H + \varepsilon \theta(t)f(t) V_\mathcal{P},
\end{equation}
where the perturbing potential $V_\mathcal{P}$ is a bounded multiplication operator $V_\mathcal{P} : \mathcal{H}_N \rightarrow \mathcal{H}_N$, the function $f \in L^\infty(\R)$, $\varepsilon \in \R$, and $\theta$ is the Heaviside function
\begin{align*}
    \theta(t) = \begin{dcases} 0, &\mbox{ if  $t<0$,} \\
    1, &\mbox{ otherwise.} \end{dcases}
\end{align*}

Next, suppose that $V_\mathcal{O}: \mathcal{H}_N \rightarrow \mathcal{H}_N$ is an observable of interest. We are interested in the expectation value $\langle V_\mathcal{O} \rangle_t \coloneqq \langle \Psi(t) , V_\mathcal{O} \Psi(t) \rangle$ for small time $t$ where $\Psi$ is the solution of the time-dependent Schr\"odinger equation
\begin{equation}
    \label{eq:TD Schrodinger1}
    \left\lbrace\begin{aligned}
        i \partial_t \Psi(t) &= H(t) \Psi(t), && t>0\\
        \Psi(0) &= \Psi_0 
    \end{aligned}\right.
\end{equation}
with $\Psi_0$ being the ground-state wave function of $H$.

%Solving the many-body Schr\"odinger equation is not feasible in practice, however it is often sufficient to have access to the first order variation of the expectation value $\langle V_\mathcal{O} \rangle_t$ with respect to $\varepsilon$. 
%We define the linear response function $O_1$ as the first order term in the expansion of $\langle V_\mathcal{O} \rangle_t$ with respect to $\varepsilon$
%\[
%\langle V_\mathcal{O} \rangle_t = \langle V_\mathcal{O} \rangle_0 + \varepsilon O_1(t) + \cO(\varepsilon^2).
%\]

\begin{prop}
\label{prop:taylor}
Let $H(t)$ be the family of self-adjoint operators defined in \eqref{eq:TD Hamiltonian}. Let $\Psi(t)$ be the solution of \eqref{eq:TD Schrodinger1} and $V_\mathcal{O} : \mathcal{H}_N\rightarrow \mathcal{H}_N$ be a bounded multiplication operator. Then $\langle V_\mathcal{O} \rangle_t = \langle \Psi(t), V_\mathcal{O}\Psi(t) \rangle$ has the following expansion
\begin{equation} \label{generalKubo}
    \langle V_\mathcal{O} \rangle_t - \langle V_\mathcal{O} \rangle_0 = i \varepsilon \int_{-\infty}^\infty  \theta(t-t') f(t')\langle {\Psi_0} , \left[ V_\mathcal{P}, (V_\mathcal{O})_I(t-t') \right]  {\Psi_0} \rangle \mathrm{d}t' + \mathcal{O}(\varepsilon^2),
\end{equation}
where $(V_\mathcal{O})_I(\tau) = e^{i\tau H} V_\mathcal{O} e^{-i\tau H}$ and $[A,B] = AB - BA$ denotes the commutator.
\end{prop}

\begin{proof}
Let $H_1(t) = \varepsilon \theta(t) f(t) V_\mathcal{P}$.
 Under the assumption on $V_\mathcal{P}$ and $f$, it is clear that $t \mapsto H_1(t) \in \mathcal{B}\bigr(\mathcal{H}_N,\mathcal{H}_N\bigr)$ is uniformly bounded. Hence, by a standard evolution equations argument, the solution to the time-dependent Schrodinger equation exists, is unique, and satisfies the Duhamel formula
\begin{equation}\label{eq:fix_point_schrodinger}
\Psi(t) = e^{-itH}\Psi_0 - i \int_0^t e^{-i(t-s)H}H_1(s) \Psi(s) \mathrm{d}s.
\end{equation}
In particular, iterating \eqref{eq:fix_point_schrodinger} twice yields
\begin{equation}\label{eq:fix_point_schrodinger2}
\Psi(t) = e^{-itE_0}\Psi_0 - i \varepsilon \int_0^t f(s)\theta(s)  e^{-i(t-s)H} V_\mathcal{P}  e^{-is H}\Psi_0 \mathrm{d}s + \mathcal{O}(\varepsilon^2).
\end{equation}
Hence plugging \eqref{eq:fix_point_schrodinger2} into the definition of $\langle V_\mathcal{O} \rangle_t$ completes the proof.
\end{proof}

If the perturbation operator $V_\mathcal{P}$ as well as the observable $V_\mathcal{O}$ are given by one-body potentials
\begin{equation}
   V_\mathcal{P} =  \sum_{k=1}^N v_\mathcal{P}(r_k), \quad V_\mathcal{O} =  \sum_{k=1}^N v_\mathcal{O}(r_k),
\end{equation}
with $v_\mathcal{O}$ and $v_\mathcal{P}$ real-valued bounded functions, then the integrand in \eqref{generalKubo} is bilinear in the potentials, providing the following
bilinear form on $L^\infty(\R^3)\times L^\infty(\R^3) $:
\begin{align}
    \label{eq:responseoperatordef}
    \langle v_\mathcal{O}, \chi(\tau) v_\mathcal{P} \rangle = i \theta(\tau) \left\langle \Psi_0 , \left[ \sum_{k=1}^N v_\mathcal{P}(r_k), \Big(\sum_{k=1}^N v_\mathcal{O}(r_k)\Big) ^{}_I(\tau) \right]  \Psi_0 \right\rangle,
\end{align}
where the inner product on the left is the $L^2$ inner product. The operator-valued function $\tau\mapsto \chi(\tau)$ defined by eq.~\eqref{eq:responseoperatordef} is the {\it density-density response function} associated with the Hamiltonian $H$.

The reason behind this definition is the celebrated Kubo formula below, which gives the first order variation of $\langle V_{\mathcal{O}} \rangle_t$ caused by $V_{\mathcal{P}}$, as a convolution against $\inner{v_{\mathcal{O}}, \chi(t) v_{\mathcal{P}}}$.

\begin{cor}[Kubo formula]
Let $\chi$ be as defined in \eqref{eq:responseoperatordef}. Then
\begin{equation}
\label{eq:linear_response_convolution}
\langle V_\mathcal{O} \rangle_t - \langle V_\mathcal{O} \rangle_0 = \varepsilon \int_0^\infty \langle v_\mathcal{O}, \chi(t-t') v_\mathcal{P} \rangle f(t') \, \mathrm{d}t' + \cO(\varepsilon^2).
\end{equation}
\end{cor}

%\textcolor{gray}{\begin{note} Observe that the linear response function depends not only on the static Hamiltonian $H$, but also on the initial state $\Psi_0$. In the next section we consider the case where $\Psi_0$ is the ground state wave function of $H$. \end{note}}

%Since the linear response involves a convolution with the retarded linear response function $\chi$, it is more convenient to write the linear response in frequency. 
%Formally by a Fourier transform in time, we obtain
%\begin{align*}
%    \mathcal{F}[\langle V_\mathcal{O} \rangle_t - \langle V_\mathcal{O} \rangle_0](\omega) = \varepsilon \mathcal{F}[\langle v_\mathcal{O}, \chi v_\mathcal{P} \rangle](\omega) \widehat{f}(\omega) + \cO(\varepsilon^2). 
%\end{align*}

%An important quantity to interpret the linear response is the \emph{density-density linear response function} (DDRF) which is defined as the kernel of the linear response $\chi(\tau)$.
%
%\todo[author=Mi-Song,inline]{Is it clear that $\chi(\tau)$ has an integral kernel? Maps $A : \rmL^p \to \rmL^\infty$ have integral kernels (see Thm A.1.1) in \cite{simon86schrodinger}. Here we have the other way around and a duality argument cannot be directly applied.}

\subsection{Regularity of the density-density response function}

We now give an alternative representation of the density-density response function of $H$ and derive a few $L^p$-mapping properties that will be useful in the next sections.

%First, note that the linear response depends on both the static Hamiltonian $H$ and the initial state (before the perturbation is turned on) $\psi_0$. Since we are interested in the linear response when $\psi_0$ is the ground state of $H$, 

\begin{prop} 
    \label{prop:chi_definition}
    Let $H$ be a Hamiltonian satisfying Assumption~\ref{assump:groundstate}. Let $E_0$ be the lowest eigenvalue of $H$ and $\Psi_0$ an associated normalized eigenfunction. Define $S : \mathcal{H}_N \rightarrow  L^1(\R^3)$ to be the mapping from a many-body wavefunction $\Phi$ to the diagonal of the mixed one-body reduced density matrix $\gamma^{\Phi,\Psi_0}$, that is to say
    \begin{equation}
      (S \Phi)(r) = N \int_{(\R^3)^{N-1}} \Phi(r,r_2,\dots,r_N) \overline{\Psi_0(r,r_2,...,r_N)} \,  \mathrm{d}r_2...\mathrm{d}r_N. \label{eq:Sdef}  
    \end{equation}  
    %\textcolor{green}{GF: I SWITCHED THE ORDER OF THE TWO FACTORS IN THE INTEGRAND, IN LINE WITH THE STANDARD DEFINITION OF MIXED RDMs.} 
    Let $\chi$ be the density-density response function of $H$ (as defined in \eqref{eq:responseoperatordef}), then 
    for any $f, g \in L^\infty(\R^3)$ real-valued, one has
    \begin{align*}
        \inner{f, \chi(t) g} = \inner{f, 2\theta(t) S \sin\bigr(t (E_0-H)\bigr) S^\ast g},
    \end{align*} 
    where $S^\ast :L^\infty(\R^3) \rightarrow \mathcal{H}_N$ is the adjoint of $S$ given by
\begin{equation}
    (S^\ast v)(r_1,..,r_N) = \sum_{k=1}^N v(r_k) \Psi_0(r_1,...,r_N). \label{eq:Sadjointdef}
\end{equation}
\end{prop}

\begin{proof} First note that the Hamiltonian $H$ commutes with complex conjugation, i.e.,
\[ \overline{H \Phi}(r_1,...,r_N) = H \overline{\Phi}(r_1,...,r_N). \]
Hence, the projection-valued (spectral) measure $P^{H}_\lambda$ associated to $H$ also commutes with complex conjugation, and therefore,
\begin{align*}
    \overline{\bigr(e^{itH} \Phi\bigr)}(r_1,...,r_N) = \bigr(e^{-itH} \Phi\bigr)(r_1,...,r_N),
\end{align*}
for any real-valued wave function $\Phi \in \mathcal{H}_N$. Moreover, due to the uniqueness assumption on the ground state of $H$, we can take $\Psi_0$ to be real-valued. Therefore, using the identity $e^{itH_N}\Psi_0 = e^{itE_0} \Psi_0$ we conclude that
\begin{align*}
    \inner{f, \chi(t) g} &= i\theta(t) \bigr(\inner{ S^\ast f, e^{it(H-E_0)} S^\ast g} - \inner{S^\ast g, e^{-it(H-E_0)} S^\ast f}\bigr) \\
    &= i\theta(t) \bigr(\inner{S^\ast f, e^{it(H-E_0)} S^\ast g} - \inner{S^\ast f, \overline{e^{it(H-E_0)}S^\ast g}}\bigr) \\
    & = -2\theta(t) \inner{S^\ast f, \sin\bigr(t(H-E_0)\bigr) S^\ast g } = \inner{f,\chi_0(t) g},
\end{align*}
for any real-valued functions $f,g \in L^\infty(\R^3)$. \end{proof}

Using the boundedness of the electronic density $\rho^{\Psi_0}$, we can show that $\chi$ has more regularity (in terms of $L^p$ spaces) than simply mapping $L^\infty$ to $L^1$.
\begin{prop}[$L^p$-regularity of $\chi$] \label{prop:Lpregularity}
Let $S$ and $S^\ast$ be defined by \eqref{eq:Sdef} and \eqref{eq:Sadjointdef}, for some $H$ satisfying Assumption~\ref{assump:groundstate}. Then, $S \in \mathcal{B}(\mathcal{H}_N,L^1(\R^3)\cap L^2(\R^3))$ and $S^\ast \in \mathcal{B}(L^2(\R^3) + L^\infty(\R^3), \mathcal{H}_N)$. In particular, $t\mapsto \chi(t)$ is a strongly continuous family of operators in $\mathcal{B}\bigr(L^2(\R^3)+L^\infty(\R^3),L^1(\R^3) \cap L^2(\R^3)\bigr)$, and the map $t \mapsto \norm{\chi(t)}_{L^2+L^\infty,L^1\cap L^2}$ is uniformly bounded in $\R_+$.
\end{prop}
\begin{proof} From the boundedness of $\rho^{\Psi_0}$ and the Cauchy-Schwarz inequality we have
\begin{align*}
    \int_{\R^3} |S \Phi(r)|^2 \mathrm{d} r &\leq N\int_{\R^3} \rho^{\Psi_0}(r) \int_{\R^{3(N-1)}} |\Phi(r,r_2,\dots,r_N)|^2 \, \mathrm{d}r_2 \dots \mathrm{d}r_N  \mathrm{d}r \leq N\norm{\rho^{\Psi_0}}_{L^\infty}\norm{\Phi}_{L^2(\R^{3N})},
\end{align*}
and
\begin{align*}
    \int_{\R^3} |S \Phi (r)|\mathrm{d} r &\leq N \int_{(\R^3)^N} |\Psi_0(r,r_2,..,r_N)\Phi(r,r_2,...,r_N)|\mathrm{d}r_2...\mathrm{d}r_N \\
     &\leq N \norm{\Psi_0}_{L^2(\R^{3N})}\norm{\Phi}_{L^2(\R^{3N})} = N \norm{\Phi}_{L^2(\R^{3N})}.
\end{align*}
Hence, $S$ maps $\mathcal{H}_N$ to $L^1(\R^3)\cap L^2(\R^3)$. Since $S^\ast$ is the adjoint of $S$, $S^\ast$ is bounded from $\bigr(L^1(\R^3)\cap L^2(\R^3)\bigr)^\ast = L^2(\R^3)+L^\infty(\R^3)$ to $\bigr(\mathcal{H}_N\bigr)^\ast = \mathcal{H}_N$ (where we use the Riesz representation for the second identification). The properties of $\chi$ now follows because $\sin\bigr(t (H-E_0)\bigr)$ is strongly continuous in $\mathcal{B}(\mathcal{H}_N,\mathcal{H}_N)$ and goes to $0$ strongly as $t \rightarrow 0$.
\end{proof}

\subsection{Fourier transform of the density-density response function and its poles}
\label{sec:Fourier}

Here we give a representation of the Fourier transform of the density-density response function $\chi$ in terms of the resolvent of $H$. This can be viewed as a  mathematically rigorous version of the celebrated Lehmann representation, to which it reduces under the -- in the physics literature tacitly made but for the physical Hamiltonian \eqref{eq:hamiltoniandef} incorrect -- assumption of purely discrete spectrum. We also introduce the definition of a meromorphic family of operators with poles of finite rank, borrowed from \cite[Appendix C]{dyatlov2019mathematical}, and show that the poles of $\widehat{\chi}$ are located at the spectrum of $H$. Let us start with the Fourier transform of $\chi$, defined on the upper half plane $\{{\rm Im}(z)>0\}$ by
\begin{align*}
    \widehat{\chi}(z) = \int_0^\infty \chi(t)e^{i z t}\mathrm{d}t,
\end{align*}
{where we have used that $\chi(t)=0$ for $t<0$.}

\begin{prop}[Fourier transform of density-density response function] \label{prop:laplace}
Let $\chi$ be the density-density response function defined in \eqref{eq:responseoperatordef} for some $H$ satisfying Assumption~\ref{assump:groundstate}. Then the Fourier transform of $\chi$ is given by
\begin{align} 
    \widehat{\chi}(z) = S (1-P^H_{E_0})\bigr((E_0-z-H)^{-1} + (E_0+z-H)^{-1}\bigr)(1-P^H_{E_0}) S^\ast \quad \mbox{for any $\mathrm{Im}(z) >0$,} \label{eq:chi0laplace}
\end{align}
where the operators $S$ and $S^\ast$ are defined in \eqref{eq:Sdef},\eqref{eq:Sadjointdef}, $E_0$ is the ground state of $H$, and  $P_{E_0}^H$ is the orthogonal projection onto the space spanned by $\Psi_0$. Moreover, the Fourier transform of $\chi$ along the real line is the tempered distribution given by
\begin{align}
    \widehat{\chi}(\omega) = \lim_{\eta \ra 0^+} S (1-P^H_{E_0}) \bigr( (E_0 - \omega - i\eta-H)^{-1} + (E_0+\omega + i \eta-H)^{-1} \bigr)(1-P^H_{E_0})S^\ast, \label{eq:chi0fourier}
\end{align}
where the limit exists in the distributional sense.
\end{prop}

\begin{proof} Since $H$ is self-adjoint, from Proposition~\ref{prop:chi_definition} and the spectral theorem we find that
\begin{align*}
    \int_0^\infty \chi(t) e^{i (\omega +i \eta)t} \mathrm{d} t
    &= S \int_0^\infty \int_{E_0}^\infty 2\sin\bigr(t(E_0-\lambda)\bigr) e^{i(\omega+i \eta)t} \mathrm{d}P^{H}_\lambda \mathrm{d} t S^\ast \\
    & = S \int_{E_0}^\infty  \frac{1}{E_0 - \omega -i \eta - \lambda}+\frac{1}{E_0+\omega + i \eta - \lambda} \mathrm{d}P^{H}_\lambda S^\ast \\
    &= S (1-P^H_{E_0}) \bigr((E_0-\omega - i \eta -H)^{-1} + (E_0+\omega+i\eta -H)^{-1}\bigr) (1-P^H_{E_0}) S^\ast,
\end{align*}
where $P^{H}_\lambda$ is the spectral projection of $H$ and we have used that $(E_0+\omega + i \eta- H)^{-1}\Psi_0 + (E_0 - \omega -i \eta-H)^{-1}\Psi_0 = 0$.
\end{proof}

Next, we make formula \eqref{eq:chi0fourier} more explicit in terms of the spectrum of $H$. 

\begin{prop}[Rigorous Lehmann representation]  \label{Prop:Lehmann1}
Under the assumptions of Proposition \ref{prop:laplace}, for $\mathrm{Im}(z)>0$ we have
\begin{multline}
    \widehat{\chi}(z) = \sum_{E_j \in \sigma_d(H)\setminus \lbrace E_0 \rbrace} S P^H_{E_j} \Big( \frac{1}{E_0-z-E_j} + \frac{1}{E_0+z-E_j} \Big) P^H_{E_j} S^* \\ + S \int_{\sigma_{\mathrm{ess}}(H)} \frac{1}{E_0-z-\lambda} + \frac{1}{E_0+z-\lambda} \, \mathrm{d}P^H_\lambda S^*, \label{pre-lehmann}
\end{multline}
where $P^H_\lambda$ is the spectral projection of $H$ and $\sigma_d$ and $\sigma_{ess}$ denote the discrete respectively essential spectrum of $H$.
\end{prop}

\begin{proof} This follows immediately from Proposition \ref{prop:laplace} and the spectral theorem for selfadjoint operators.
\end{proof}

\begin{note}[Lehmann representation in the physics literature] \label{DDRF}
When $H$ has purely discrete spectrum, as happens e.g. when $v$ is a trapping potential, formula (2.16) simplifies further. Let $\{\Psi_j\}_{j=0}^{\infty}$ be an orthonormal basis consisting of eigenstates of $H$, corresponding to eigenvalues $E_j$, and use that
$$
   (1 - P_{E_0}^H) \Phi = \sum_{j=1}^\infty \Psi_j \langle \Psi_j, \, \Phi\rangle.
$$
Introduce the excitation frequencies $\omega_j=E_j-E_0$, and calculate 
\begin{align*}
	& \langle v_\mathcal{O}, \widehat{\chi}(z) v_\mathcal{P}\rangle  \\ 
    &= \sum_{j\ge 1} \int v_\mathcal{O}(r_1) \Bigl( \widehat{\chi}(z)v_\mathcal{P}\Bigr)(r_1)\, \mathrm{d}r_1 \\
    &= \sum_{j\ge 1} N\int v_\mathcal{O}(r_1) \int \overline{\Psi_0(r_1,...,r_N)}\Psi_j(r_1,...,r_N) \mathrm{d}r_2 ... \mathrm{d}r_N \mathrm{d}r_1 \Bigl( \frac{1}{-z-\omega_j} + \frac{1}{z-\omega_j} \Bigr)
    \bigl\langle \Psi_j, S^* v_p \bigr\rangle 
\end{align*} 
and therefore, decomposing $z$ into its real and imaginary part, i.e. $z=\omega+i\eta$ ($\eta>0$), 
\begin{equation} \label{lehmann1}
\langle v_\mathcal{O}, \widehat{\chi}(\omega + i\eta) v_\mathcal{P}\rangle = 
       \sum_{j\ge 1} \bigl\langle \Psi_0, V_\mathcal{O}\Psi_j\bigr\rangle \, 
       \bigl\langle \Psi_j, V_\mathcal{P}\Psi_0\bigr\rangle \,  \Bigl( \frac{1}{-(\omega+i\eta)-\omega_j} + \frac{1}{(\omega+i\eta)-\omega_j} \Bigr).
\end{equation}
This is precisely the \textit{Lehmann representation} of the density-density response function familiar from the physics literature. This representation beautifully reveals how a frequency-dependent perturbation couples to the excitation spectrum of the system.

Note, however, that this representation, unlike ours above (eq.~\eqref{pre-lehmann}), is not strictly speaking applicable to the standard atomic and molecular Hamiltonians (\eqref{eq:hamiltoniandef} with $v(r)=-\sum_{\alpha=1}^M Z_\alpha/|r-R_\alpha|$) which contain continuous spectrum, as it misses the integral term in \eqref{pre-lehmann}.
\end{note}

Let us now recall the definition of a meromorphic family of operators as defined in  \cite[Appendix C]{dyatlov2019mathematical}.

\begin{definition}[Meromorphic family of operators]\label{def:meromorphicfamily} Let $\mathcal{D} \subset \C$ be an open set and $E,F$ be Banach spaces. We say that $K: \mathcal{D} \rightarrow \mathcal{B}(E,F)$ is a meromorphic family of operators if in a neighborhood of any $z_0 \in \mathcal{D}$, there exist finite rank operators $K_{-j} \in \mathcal{B}(E,F)$, for $1\leq j \leq k$, such that
\begin{align*}
    K(z) = K_0(z) + \sum_{j=1}^k \frac{K_{-j}}{(z-z_0)^j},
\end{align*}
where $K_0(z)$ is holomorphic near $z_0$. If $k=1$, we say that $z_0$ is a simple pole and define its rank as $\rank_{z_0}(K) = \rank K_{-1}$.
\end{definition}
Then we can relate the definition above with the representation in Proposition~\ref{prop:laplace}.

\begin{prop}[Poles of $\widehat{\chi}$]\label{prop:poleschi0} Let $\chi$ be the density-density response function defined in Proposition~\ref{prop:chi_definition} for some Hamiltonian $H$ satisfying Assumption~\ref{assump:groundstate}. Let $D_\Omega \subset \mathbb{C}$ be the set 
\begin{align}
    \mathcal{D}_\Omega = \C \setminus \bigr((-\infty,-\Omega]\cup[\Omega,+\infty)\bigr). \label{eq:domaindef}
\end{align}
Then $\widehat{\chi}:\mathcal{D}_\Omega \rightarrow \mathcal{B}(L^2+L^\infty,L^1\cap L^2)$ is a meromorphic family of operators with simple poles contained in $(-\Omega,\Omega)$. Moreover, the set of poles of $\widehat{\chi}$ is
\begin{align}
    \mathcal{P}(\widehat{\chi}) = \{ \omega \in \R : E_0 +|\omega| \in \sigma(H)\setminus E_0 \mbox{ and }  S P^H_{E_0 + |\omega|} \neq 0 \}, \label{eq:poleschi0}
\end{align}
and the rank of a pole $\omega \in \mathcal{P}(\widehat{\chi})$ is given by
\begin{align}
    \rank_\omega(\widehat{\chi}) = \rank S P^H_{E_0+|\omega|}, \label{eq:rankchipoles}
\end{align}
where $P^H_{E_0 +|\omega|}$ is the spectral projection of $H$ onto the eigenspace $\ker (H-E_0-|\omega|)$.

%\begin{enumerate}[label=(\roman*)]
%\item \label{it:meromorphicpropchi0}(Meromorphic) 
%\item \label{it:realsimplepolespropchi0} (Real simple poles) Each pole $\omega \in \mathcal{P}(\widehat{\chi})$ is simple (i.e., has order $1$) and has rank equal to the dimension of the range of $S P_{E_0+|\omega|}$.
%\item \label{it:poleschi0propchi0}(Poles of $\widehat{\chi}$) The set of poles of $\widehat{\chi}$ is given by
%\begin{align}
%    \mathcal{P}(\widehat{\chi}) = \{ \omega \in \R : E_0 +|\omega| \in \sigma(H)\setminus E_0 \mbox{ and }  S P_{E_0 + |\omega|} \neq 0 \}, \label{eq:poleschi0}
%\end{align}
%where $P_{E_0 +|\omega|}$ is the spectral projection of $H$ at the eigenspace $\ker (H-E_0-|\omega|)$.
%\end{enumerate}
\end{prop}

\begin{proof} 
For $\mathrm{Im}(z)>0$, we start from the representation of $\widehat{\chi}(z)$ in  Proposition~\ref{Prop:Lehmann1}. As $\Omega>0$, we can extend $\widehat{\chi}(z)$ analytically to the lower half plane $\mathrm{Im}(z)<0$ and we directly obtain that it is a meromorphic family of operators with simple real poles
with the characterization~\eqref{eq:poleschi0}. {For the statement on the rank of the poles, note that from \eqref{pre-lehmann}, $\rank_\omega(\widehat{\chi}) = \rank(S P^H_{E_0+\omega} S^\ast)$. Moreover
\begin{align*}
    \inner{f, S P^H_{E_0+\omega} S^\ast f} = \inner{P^H_{E_0+\omega} S^\ast f ,  P^H_{E_0+\omega} S^\ast f } = \norm{P^H_{E_0+\omega}S^\ast f}_{L^2(\R^{3N})}^2, 
\end{align*}
for any $f \in L^2(\R^3)+L^\infty(\R^3)$, and so $\rank(S P^H_{E_0+\omega} S^\ast)\geq \rank(P^H_{E_0+\omega}S^*)$.} 
But $\rank (P^H_{E_0+\omega}S^*) = \rank (S P^H_{E_0+\omega})$ and we already have $\rank(S P^H_{E_0+\omega} S^\ast) \leq \rank(S P^H_{E_0+\omega})$, hence $\rank_\omega(\widehat{\chi}) = \rank(S P^H_{E_0+\omega})$.
%and
%\begin{align*}
%    \inner{\Psi, P_{E_0+\omega}^H S^\ast S P^H_{E_0+\omega} \Psi}_{L^2(\R^{3N})} = \norm{S P^H_{E_0+\omega} \Psi}_{L^2(\R^3)}^3,
%\end{align*}
%for any $\Psi \in L^2(\R^{3N})$. It thus follows that $\rank_\omega(\widehat{\chi}) = \rank(S P_{E_0+\omega} S^\ast) = \rank(SP_{E_0+\omega})$.
%The first two items of the proposition follow from the spectral decomposition
%\begin{align*}
%    \widehat{\chi}(z) = \underbrace{S P_{\textnormal{ess}}\bigr(R(E_0+z)+R(E_0-z)\bigr)S^\ast }_{=B_{\textnormal{ess}}(z)} +  \sum_{\omega_j \in S_{\Omega}(\chi)} \frac{\textnormal{sign}(\omega_j)}{z - \omega_j}\underbrace{S P_j S^\ast}_{=B_j},
%\end{align*}
%where $P_{\textnormal{ess}}$ is the spectral projection in the interval $[\Omega,+\infty)$ and $P_j$ is the orthogonal projection at the eigenspace $\ker(H-E_0-|\omega_j|)$. Note that the condition $S P_{E_0+|\omega_j|}$ is necessary for the poles, as otherwise, the operator $B_j$ is identically zero. The last statement follows since $B_j =  (S P_j) (S P_j)^\ast$, which is injective on the range of $SP_j$ by the identity $\ker(P_jS^\ast) = \textnormal{range}(SP_j)^\ast$. 
\end{proof}

\begin{note} If the Hamiltonian $H$ has purely discrete spectrum (for instance when $v$ is a trapping potential), $\mathcal{D}_\Omega  = \C$ and $\mathcal{P}(\widehat{\chi})$ is the whole set of singular points of the Fourier transform $\widehat{\chi}$. However%,for typical applications in electronic structure however (for instance, atomic or molecular Hamiltonians) the spectrum consists of a discrete part below the bottom of the essential spectrum and a continuous part above it. In this case}
, past the ionization threshold (see \eqref{eq:ionization_threshold}) it is not clear how singular $\widehat{\chi}$ is. For instance, under suitable assumptions on $v$ and $w$ one can use the celebrated \textit{limiting absorption principle} \cite{Tamura1989,Georgescupaper,commutatormethodsbook} to show that $\widehat{\chi}$ is continuous -- or even differentiable -- above the ionization threshold (and away from embedded eigenvalues).
\end{note}

\section{The RPA Dyson equation}
\label{sec:rpa_in_time}

The goal of this section is to prove Theorem~\ref{thm:time_rpa_solution}. We start with existence and uniqueness, and then prove the bijection property. To shorten the notation, for any $T>0$ we define
\begin{align} \label{eq:Tnorm}
    \norm{\chi}_{T} \coloneqq \esssup_{t \in (0,T]} \norm{\chi(t)}_{L^2+L^\infty,L^1\cap L^2}, 
\end{align}
where $\chi \in L^\infty\bigr([0,T);\mathcal{B}(L^2+L^\infty,L^1\cap L^2)\bigr)$.

\subsection{Well-posedness of the Dyson equation}

We now turn to the well-posedness of the RPA-Dyson equation 
\begin{equation} \label{eq:dyson_general}
    \chi^\mathrm{RPA}(t) = \chi_0(t) + \int_0^t \chi_0(t-s) F_{H} \chi^\mathrm{RPA}(s) \mathrm{d} s,
\end{equation}
where $F_H$ is the Hartree operator defined as the convolution with $\frac{1}{|\cdot|}$. Here the reference response function $\chi_0$ can be a general operator-valued function of time, only required to satisfy mild regularity conditions. 

The starting point is to show that the convolution map
\begin{align}
    (\chi_0,\chi) \mapsto \mathcal{C}(\chi_0,\chi)(t) = \bigr(\chi_0 \star F_H \chi\bigr)(t) \coloneqq \int_0^t \chi_0(t-s) F_H \chi(s) \mathrm{d} s
\end{align}
is continuous in appropriate spaces. More precisely, we have the following lemma. 

\begin{lem}[Continuity of convolution map] \label{lem:hls lemma}
Let $\chi \in L^\infty((0,T],\mathcal{B}(L^{2}+L^{\infty},L^1\cap L^2))$ and $\chi_0  \in L^\infty((0,T],\mathcal{B}(L^{2}+L^{\infty},L^1\cap L^2))$. Then, the function $t\mapsto  \chi_0(t-s) F_H \chi(s)$ belongs to $L^\infty((0,T],\mathcal{B}(L^{2}+L^{\infty},L^1\cap L^2))$ and it holds 
\begin{align}
    \norm{\mathcal{C}(\chi_0,\chi)}_{T} \lesssim T \norm{\chi_0}_{T}\norm{\chi}_{T} \label{eq:contraction}
\end{align}
Moreover, if either $\chi$ or $\chi_0$ is strongly continuous, then so is $\mathcal{C}(\chi_0,\chi)$.
\end{lem}

\begin{proof} Since $F_H = 4\pi (-\tfrac{1}{2}\Delta)^{-1}$ and we have the continuous inclusions $L^{p}(\mathbb{R}^3) \subset L^2(\mathbb{R}^3) + L^\infty(\mathbb{R}^3)$ and $L^1(\mathbb{R}^3) \cap L^2(\mathbb{R}^3) \subset L^q(\mathbb{R}^3)$, for $2\leq p \leq \infty$ and $1\leq q \leq 2$, estimate~\eqref{eq:contraction} follows directly from the Hardy-Littlewood-Sobolev inequality, which in $\R^3$ reads
\[
||I_\alpha f||_{p} \leq C ||f||_{q}, \qquad I_\alpha = \big(- \Delta)^{- \nicefrac{\alpha}{2}},
\]
for $\tfrac{1}{q} = \tfrac{1}{p} + \tfrac{\alpha}{3}$ with $1<p,q<\infty$. The strong continuity follows by observing that $\chi_0(t-s) F_H \chi(s)$ is strongly continuous in $t$ and uniformly bounded (in $s$) in the $\mathcal{B}(L^2 +L^\infty,L^1 \cap L^2)$-operator norm. Hence, by dominated convergence we find that $\mathcal{C}(\chi_0,\chi)$ is also strongly continuous. On the other hand, if $\chi$ is strongly continuous we can use the change of variables $s\mapsto t-s$ and the same argument to show that $\mathcal{C}(\chi_0,\chi)$ is strongly continuous. %one has
%\begin{align*}
%    \mathcal{C}(\chi_0,\chi)(t) = \int_0^t \chi_0(s) F_H \chi(t-s) \mathrm{d}s
%\end{align*}
%Hence, the integrand is again bounded and strongly continuous in $t$, thus, the result follows again by dominated convergence.
\end{proof}
We can now use the above estimate to show the well-posedness on the space of strongly continuous $\mathcal{B}(L^2+L^\infty,L^1\cap L^2)$-valued functions.
\begin{proof}[Proof of item~\ref{itthm:existence_rpa} from Theorem~\ref{thm:time_rpa_solution}]
First note that by inequality \eqref{eq:contraction}, for $T$ sufficiently small we know that the map
\begin{equation*}
    \mathcal{C}^T(\chi_0,\cdot) \ : \ \left\lbrace
    \begin{aligned} L^\infty((0,T],\mathcal{B}(L^2+L^\infty,L^1\cap L^2))&\rightarrow L^\infty((0,T],\mathcal{B}(L^2+L^\infty,L^1\cap L^2)) \\
    \chi &\mapsto \mathcal{C}(\chi_0,\chi)(T) 
    \end{aligned}\right.
\end{equation*}
is a contraction. Therefore, by the Banach fixed point theorem, there exists a unique solution $\chi_1 \in L^\infty\bigr((0,T];\mathcal{B}(L^2+L^\infty,L^1\cap L^2)\bigr)$ satisfying $\chi_1 = \chi_0 + \mathcal{C}^T(\chi_0,\chi_1)$. Hence, we just need to extend this solution to the whole $\R_+$. For this, note that, as $\chi_0 \in L^\infty((0,2T],\mathcal{B}(L^2+L^\infty,L^1\cap L^2))$, there exists some {$\delta  = \delta(\norm{\chi_0}_{{2T}}) >0$ (with the norm defined in \eqref{eq:Tnorm})} such that the map 
\begin{align*}
    \mathcal{C}^{T+\delta}_{T_0}(\chi_0,\chi)(t) \coloneqq \int_{T_0}^{t}\chi_0(t-s)F_H \chi(s) \mathrm{d} s
\end{align*}
is also a contraction in $L^\infty([T_0-\delta,T_0+\delta];\mathcal{B}(L^2+L^\infty,L^1\cap L^2))$ for any $0 <T_0 \leq 2T-\delta$. Hence, let $\chi_1(t)$ be the solution in $L^\infty\bigr((0,T];\mathcal{B}(L^2+L^\infty,L^1\cap L^2)\bigr)$, then the map
\begin{align*} \chi \mapsto \chi_0 + \int_0^{T} \chi_0(t-s) F_H \chi_1(s) \mathrm{d} s + \mathcal{C}^{T+\delta}_{T}(\chi_0,\chi)  \end{align*}
is again a contraction and we can find a unique fixed point $\chi_2$. Moreover, we have
\begin{align*} \chi_1(t) - \chi_2(t) = \int_{T}^t \chi_0(t-s) F_H \bigr(\chi_1(s) - \chi_2(s)\bigr) \mathrm{d} s = \mathcal{C}_{T}^{T+\delta} (\chi_0,\chi_1-\chi_2)(t),\end{align*}
for any  $T-\delta < t \leq T+\delta$. But because $0$ is the unique fixed point of $\mathcal{C}^{T+\delta}_T(\chi_0, \cdot)$, we must have $\chi_1(t) = \chi_2(t)$ for a.e $T-\delta < t\leq T + \delta$. We have thus extended $\chi_1$ to $(0,T+\delta]$. To conclude, note that since $\delta$ is uniform in the interval $(0,2T]$, we can iterate the argument to extend $\chi_1$ to the interval $(0,2T]$. Repeating the same steps, we can further extend the solution to the whole $\R_+$. The strong continuity follows from the strong continuity in Lemma~\ref{lem:hls lemma}.
\end{proof}

\subsection{Bijection of the RPA-Dyson solution map}

In virtue of Theorem~\ref{thm:time_rpa_solution}~\ref{itthm:existence_rpa}, we can define the solution map
\begin{equation*} 
    \mathcal{S}^{RPA} \ : \ \left\lbrace
    \begin{aligned}
        C_s\bigr(\mathbb{R}_+;\mathcal{B}(L^2+L^\infty,L^1\cap L^2)\bigr) &\rightarrow C_s\bigr(\mathbb{R}_+;\mathcal{B}(L^2+L^\infty,L^1\cap L^2)\bigr)\\
        \chi_0  &\mapsto \chi^\mathrm{RPA} \in \ker\{\chi_0 + \mathcal{C}(\chi_0,\cdot)-\cdot\}.  
    \end{aligned}\right.
\end{equation*}
To complete the proof of Theorem~\ref{thm:time_rpa_solution}, we now show that $\mathcal{S}^{RPA}$ is bijective in $C_s\bigr(\mathbb{R}_+;\mathcal{B}(L^2+L^\infty,L^1\cap L^2)\bigr)$. 
\begin{proof}[Proof of item~\ref{itthm:bijection_rpa} of Theorem~\ref{thm:time_rpa_solution}] Note that, by repeating the arguments in the proof of item~\ref{itthm:existence_rpa} of Theorem~\ref{thm:time_rpa_solution}, for any $\chi \in C_s\bigr(\mathbb{R}_+;\mathcal{B}(L^2+L^\infty,L^1\cap L^2)\bigr)$, we can find a unique $\chi_0 \in C_s\bigr(\mathbb{R}_+;\mathcal{B}(L^2+L^\infty,L^1\cap L^2)\bigr)$ satisfying $\chi_0 = \chi - \mathcal{C}(\chi_0,\chi)$. In particular, $\chi = \mathcal{S}^{RPA}(\chi_0)$ is the unique solution of the RPA-Dyson equation, which implies that $\mathcal{S}^{RPA}$ is surjective. Similarly, by the uniqueness of the solution $\chi_0$ of $\chi_0 = \chi - \mathcal{C}(\chi_0,\chi)$, we also have injectivity of $\mathcal{S}^{RPA}$ and the proof is complete. \end{proof}
%\begin{note} Note that the only property of $F_H$ we needed is that $F_H \in \mathcal{B}(\rmL^q,\rmL^p)$ for some $p \in [2,\infty]$ and $q \in [1,2]$. In particular, one can use the same arguments here to prove the well-posedness of the Dyson equation for other adiabatic approximations of the exchange correlation kernel. \end{note}}

\section{Symmetrized density-density response function}
\label{sec:symmetrized_chi}

In this section, we want to characterize the poles of $(1-\widehat{\chi}_s(z))^{-1}$ for operators of the form
\begin{align}
    \widehat{\chi}_s(z) = T \biggr(\int_{\sigma(H)} 2 \frac{E_0-\lambda}{(E_0-\lambda)^2 - z^2} \mathrm{d} P^{H}_\lambda \biggr) T^\ast, \label{eq:def_symmetrized_chi}
\end{align}
where $T$ is a bounded operator from $\mathcal{H}_N$ to a Hilbert space $\mathcal{H}$, and $P^{H}_\lambda$ is the projection-valued measure of a Hamiltonian satisfying Assumption~\ref{assump:groundstate}. {As already remarked at the end of the Introduction, such a characterization is essential for the proof of Theorem~\ref{theo:polesRPA}.} 

We start by introducing some new notation. In virtue of Proposition~\ref{prop:poleschi0}, we define the relevant excitations $0 < \omega_1 < \omega_2 < \dots $ as the set of positive poles of $\widehat{\chi}_s$, i.e., 
\begin{align}
    \{\omega_j\}_{j=1}^m = \{0<\omega< \Omega : T P^H_{E_0+\omega} \neq 0\} = \mathcal{P}(\widehat{\chi}_s) \cap (0,\Omega) \label{eq:excitationdef}
\end{align}
where $P^H_{E_0 + \omega}$ is the spectral projection of $H$ at the eigenvalue $E_0 + \omega$.
For finite $m$, we set $\omega_{m+1} = \Omega$, where we recall that $\Omega$ is the ionization threshold defined in \eqref{eq:ionization_threshold}.
We also call the excitation-free intervals, the intervals $(\omega_j,\omega_{j+1})$, and define the finite-dimensional subspaces
\begin{align}
    V_j \coloneqq (\ker P^H_{E_0+\omega_j} T^\ast)^\perp = \ran TP^H_{E_0+\omega_j}. \label{eq:Vjdef}
\end{align}
Then by Proposition~\ref{prop:poleschi0} (which also holds for $\widehat{\chi}_s$ in the place of $\widehat{\chi}$), the rank of $\omega_j$ as a pole of $\widehat{\chi}_s$ is given by $\dim V_j$.
\begin{note}\label{note:Tker} Note that we can assume $\ker (1-P^H_{E_0})T^\ast = \{0\}$, as otherwise, we could simply set $\widetilde{\mathcal{H}} = (\ker (1-P^H_{E_0}) T^\ast)^\perp = \overline{\ran(T (1-P^H_{E_0}))} \subset \mathcal{H}$ and consider $\widehat{\chi}_s = P_{\widetilde{\mathcal{H}}} \widehat{\chi}_s P_{\widetilde{\mathcal{H}}}$ as an operator in $\widetilde{\mathcal{H}}$. \end{note}
The main goal of this section is then to prove the following propositions.
\begin{prop}[Characterization of the poles of $(1-\widehat{\chi}_s)^{-1}$]\label{prop:1-chi_s_meromorphic} Let $\widehat{\chi}_s(z)$ be defined by \eqref{eq:def_symmetrized_chi} and $\mathcal{D}_\Omega = \C \setminus \big( (-\infty,-\Omega] \cup [\Omega,\infty) \big)$. Then $(1-\widehat{\chi}_s)^{-1} : \mathcal{D}_\Omega \mapsto \mathcal{B}(\mathcal{H},\mathcal{H})$  is a meromorphic family of operators with simple real poles with rank given by
\begin{align}
    \rank_{\omega}\bigr((1-\widehat{\chi}_s)^{-1}\bigr) = \begin{dcases} \dim\ker\bigr( 1 - P_{V_j}^\perp \widehat{\chi}_s(\omega_j) P_{V_j}^\perp \bigr) , &\mbox{ if $\omega = \omega_j$ for some $j \leq m$,}\\
    \dim \ker \bigr(1-\widehat{\chi}_s(\omega)\bigr), &\mbox{otherwise.}\end{dcases} \label{eq:rankpoleschis}
\end{align}
Moreover, for $z$ close to $\omega_j$ we have
\begin{align}
    \bigr(1-\widehat{\chi}_s(z)\bigr)^{-1} = (z-\omega_j)^{-1} K_{-1} + K_0 + \mathcal{O}(|z-\omega_j|) , \label{eq:detailedpoleinfo}
\end{align}
where $K_{-1} P_{V_j} =0$ and $\ran K_0 P_{V_j} \subset \ran K_{-1}$.
\end{prop}
\begin{prop}[Forward shift of the poles of $(1-\widehat{\chi}_s)^{-1}$]\label{prop:forwardshiftchis}
    Let $\widehat{\chi}_s$ be the operator defined in \eqref{eq:def_symmetrized_chi}.
 Then, the poles of $(1-\widehat{\chi}_s)^{-1}$ are forward shifted with respect to the poles of $\widehat{\chi}_s$ in the sense that for any $0<\omega < \Omega$, we have
\begin{align}
    \sum_{\substack{\widetilde{\omega} \in \mathcal{P}( (1-\widehat{\chi}_s)^{-1}) \\ |\widetilde{\omega}|<\omega}} \rank_{\widetilde{\omega}}\bigr((1-\widehat{\chi}_s)^{-1}\bigr) \leq \sum_{\substack{\widetilde{\omega} \in \mathcal{P}(\widehat{\chi}_s) \\ |\widetilde{\omega}|<\omega}} \rank_{\widetilde{\omega}}(\widehat{\chi}_s). \label{eq:forwardshift}
\end{align}
\end{prop}
{A very crude heuristics for why such a statement might be true goes as follows. Assume that $\widehat{\chi}_s$ equals $\frac{a}{z-\omega}$ with $a\neq 0$, which is the simplest function exhibiting a pole at $\omega$. Then $(1-\widehat{\chi}_s(z))^{-1}=\frac{z-\omega}{z-(\omega + a)}$, which has a pole at $\omega+a$. Thus when $a>0$, the pole of $(1-\widehat{\chi}_s)^{-1}$ is forward-shifted with respect to that of $\widehat{\chi}_s$. 
} % end blue

The plan for the rest of the section is to first prove Proposition~\ref{prop:1-chi_s_meromorphic} and then use it to prove Proposition~\ref{prop:forwardshiftchis}. The proof of Proposition~\ref{prop:1-chi_s_meromorphic} consists of three main steps. First we show that $1-\widehat{\chi}_s(z)$ is invertible for $\mathrm{Im}(z) \neq 0$ or $|\omega| < \omega_1$ and find an explicit estimate for the blow-up of the inverse as $\mathrm{Im}(z) \ra 0$ (Lemma~\ref{lem:inverseawayofimaginary}). We then use this estimate to conclude that all poles in the excitation-free intervals $(\omega_j,\omega_{j+1})$ are simple and give their rank (Lemma~\ref{lem:simple_poles_normal_op}). Finally we isolate the singularity of $1-\widehat{\chi}_s(\omega)$ at $\omega = \omega_j$ and deal with the blowing-up and vanishing part of $1-\widehat{\chi}_s(\omega)$ separately. 

\subsection{Proof of Proposition~\ref{prop:1-chi_s_meromorphic}}

The starting point of our analysis is the spectral decomposition 
\begin{align}
    \widehat{\chi}_s(z) = \sum_{j\leq m} \frac{2\omega_j}{z^2-\omega_j^2} \underbrace{T P^H_{E_0+\omega_j} T^\ast}_{\coloneqq B_j} + \underbrace{T \biggr(\int_{\Omega}^{{\infty}} \frac{2\lambda}{z^2-\lambda^2}\mathrm{d} P^{H}_{E_0+\lambda}\biggr) T^\ast}_{\coloneqq B_{\textnormal{ess}}(z)}, \label{eq:maindecomposition}
\end{align}
where $P^H_{E_0 + \omega_j}$ is the spectral projection in the eigenspace $\ker (H-E_0-|\omega_j|)$, and  $\omega_j$ are the excitations defined in \eqref{eq:excitationdef}. Note that since $T P^H_{E_0+\omega_j} T^\ast =  T P^H_{E_0+\omega_j} (T P^H_{E_0+\omega_j})^\ast$, the operators $B_j$ are invertible with respect to the orthogonal projection $P_{V_j}$. This follows from the identity $\ker A = (\ran A)^\perp$ valid for any symmetric operator $A \in \mathcal{B}(\mathcal{H},\mathcal{H})$.

The first step to prove Proposition~\ref{prop:1-chi_s_meromorphic} is to show that the positive spectra of $\widehat{\chi}_s$ is discrete. This follows from the following proposition.
\begin{prop}[Essential spectrum of  $\widehat{\chi}_s$]
    \label{prop:essentialspec}
    Let $\widehat{\chi}_s(z)$ be defined by \eqref{eq:def_symmetrized_chi}. Then, for any $z \in \mathcal{D}_\Omega \setminus \mathcal{P}(\widehat{\chi}_s)$, the operator $\widehat{\chi}_s(z)$ satisfies
    \begin{align}
        \widehat{\chi}_s(z) = \widehat{\chi}_s(\overline{z})^\ast = \widehat{\chi}_s(-z). \label{eq:symmetrieschi}
    \end{align}
    In particular, $\widehat{\chi}_s(z)$ is self-adjoint for real $z$. Furthermore, we have
    \begin{align}
      \sigma_{\textnormal{ess}}\bigr(\widehat{\chi}_s(\omega)\bigr) \subset (-\infty,0], \label{eq:essentialnonpositive}
    \end{align}
    for any $\omega \in (-\Omega,\Omega)\setminus \mathcal{P}(\widehat{\chi}_s)$.
\end{prop}
\begin{proof}
The symmetries in \eqref{eq:symmetrieschi} are immediate from the definition in \eqref{eq:def_symmetrized_chi} and the identity $\big((z-H)^{-1}\big)^\ast = (\overline{z}-H)^{-1}$. For the essential spectrum part, note that $\frac{2\lambda}{\omega^2-\lambda^2} < 0$ for $|\omega|< \lambda$. This together with the fact that $B_j = (T P^H_{E_0+\omega_j}) (T P^H_{E_0+\omega_j})^\ast$ is non-negative implies that
\begin{align}
    &\inner{f, \frac{2\omega_j}{\omega_j^2-\lambda^2} B_j f} \leq 0 \quad \mbox{ for any $0<\omega < \omega_j$, and} \label{eq:nonpositive1} \\
    &\inner{f, B_{\textnormal{ess}}(\omega) f} \leq 0 \quad \mbox{ for any $0<\omega <\Omega$.} \label{eq:nonpositive2}
\end{align}
In addition, since all $B_j$'s are finite rank operators, from Weyl's criterion we have
\begin{align}
    \sigma_{\textnormal{ess}}\bigr(\widehat{\chi}_s(\omega)\bigr) = \sigma_{\textnormal{ess}}\biggr( \sum_{j \geq k} \frac{2\omega_j}{\omega^2-\omega_j^2} B_j + B_{\textnormal{ess}}(\omega)\biggr) \quad \mbox{ for any integer $k \leq  m$ . } \label{eq:essentialnonpositiveest}
\end{align}
%Thus since any $\omega \in (0,\Omega)\setminus \mathcal{P}(\widehat{\chi}_s)$ must belong to one of the excitation-free intervals $(\omega_j,\omega_{j+1})$, 
The result now follows from \eqref{eq:nonpositive1}, \eqref{eq:nonpositive2} and \eqref{eq:essentialnonpositiveest} by the Rayleigh Ritz principle.
\end{proof}

\subsubsection{Inverse of $1-\widehat{\chi}_s(z)$ for $\mathrm{Im} (z) \not=0$ or $|z| < \omega_1$ .}

Next, we want to show that $1-\widehat{\chi}_s(z)$ is invertible for any $z$ with $\mathrm{Im}(z) \neq 0$ or $|\mathrm{Re}(z)|<\omega_1$. For this, we shall use the following inequality between the real and imaginary part of $\inner{f,\widehat{\chi}_s(z)f}$.
\begin{lem}[Real to imaginary ratio] \label{lem:real_to_imaginary_ratio} Let $\widehat{\chi}_s(z)$ be defined by \eqref{eq:def_symmetrized_chi}, then
for any $z {= \omega+ i \eta} \in \mathcal{D}_\Omega\setminus\bigr( (-\Omega, - \omega_1] \cup [\omega_1, \Omega)\bigr)$ and $f \in \mathcal{H}$ we have
\begin{align}
    \mathrm{Re}\bigr(\inner{f,\widehat{\chi}_s(z)f}\bigr) \leq \max \biggr\{ 0, \frac{\omega^2 - \eta^2 - \omega_1^2}{|\omega \eta|}\biggr\} \bigr|\mathrm{Im}\bigr(\inner{f,\widehat{\chi}_s(z) f}\bigr)\bigr|. \label{eq:realtoimaginary}
\end{align}
\end{lem}

\begin{proof} 
Let $z = \omega + i \eta$ and suppose that $\omega^2-\eta^2 \leq \omega_1^2$. Then since the integrand in \eqref{eq:def_symmetrized_chi} vanishes for $\lambda = E_0$, by making the translation $\lambda \mapsto \lambda-E_0$, we find that
\begin{align*}
    \mathrm{Re} \inner{f,\widehat{\chi}_s(z) f} = 2\int_{\omega_1}^\infty \frac{\overbrace{\lambda(\omega^2-\eta^2-\lambda^2)}^{\leq 0}}{|\lambda^2 + z^2|^2} \mathrm{d} \norm{P^{H}_{E_0+\lambda }T f}^2 \leq 0
\end{align*}
which gives estimate \eqref{eq:realtoimaginary} in this case. On the other hand, if $\omega^2-\eta^2>\omega_1^2$ we have %, since the integrand is negative for $\lambda \geq \sqrt{\omega^2-\eta^2}$, one has
\begin{align*}
    \mathrm{Re} \inner{f, \widehat{\chi}_s(z) f} &\leq 2\int_{\omega_1}^{ \sqrt{\omega^2-\eta^2}}\frac{\lambda(\omega^2-\eta^2-\lambda^2)}{|\lambda^2 - z^2|^2} \mathrm{d} \norm{P^{H}_{E_0+\lambda} T f}^2 \\
    &\leq 2\int_{\omega_1}^{ \sqrt{\omega^2-\eta^2}} \frac{\lambda |\omega \eta|}{|\lambda^2-z^2|^2} \biggr(\frac{\omega^2-\eta^2-\lambda^2}{|\omega \eta|}\biggr) \mathrm{d}\norm{P^{H}_{E_0+\lambda} Tf}^2 \\
    &\leq \frac{\omega^2-\eta^2-\omega_1^2}{|\omega \eta|}\biggr|2 \int_{\omega_1}^{ \sqrt{\omega^2-\eta^2}} \frac{\lambda \omega \eta }{|\lambda^2-z^2|^2}  \mathrm{d}\norm{P^{H}_{E_0+\lambda} T f}^2 \biggr|\\
    &\leq \frac{\omega^2-\eta^2-\omega_1^2}{|\omega \eta|} |\mathrm{Im}\inner{f, \widehat{\chi}_s(z) f}| .
\end{align*}\end{proof}

Now we can use estimate~\eqref{eq:realtoimaginary} to show that $1-\widehat{\chi}_s(z)$ is invertible away from the real axis and before the first excitation $\omega_1$. In addition, we obtain an explicit upper bound on the blow-up rate of the inverse as $z$ approaches the real axis. This bound will be useful to show that the poles of $(1-\widehat{\chi}_s)^{-1}$ are simple.
\begin{lem}[Inverse away of the real axis and before $\omega_1$]\label{lem:inverseawayofimaginary} Let $\widehat{\chi}_s(z)$ be defined in \eqref{eq:def_symmetrized_chi} and $\mu_0>0$. Then $\mu_0 - \widehat{\chi}_s(z)$ is invertible in the set $\{ z = \omega + i \eta \in \C: \eta \neq 0 \mbox{ or } |\omega| \leq \omega_1\}$. Moreover, we have
\begin{align}
    \norm{(\mu_0 - \widehat{\chi}_s(\omega + i \eta ))^{-1}} \lesssim  \mu_0^{-1} |z| |\eta|^{-1},  \label{eq:inverseblowuprate}
\end{align}
for any $z\in \C\setminus \R$.
\end{lem}

\begin{proof} Let $g \, : \, \mathbb{C} \setminus \{\omega_1, - \omega_1\} \to \mathbb{R} \cup \{+\infty\}$  be the function
\begin{align}
    g(\omega + i \eta) = \max \biggr\{ 0, \frac{\omega^2 - \eta^2 - \omega_1^2}{|\omega \eta|}\biggr\}, \quad (\omega, \eta) \in \mathbb{R}^2 \setminus \{(\omega_1,0),(0,\omega_1)\}.
\end{align}
Then, for $f\neq 0$ with $\mathrm{Re} \inner{f, \widehat{\chi}_s(z) f} \leq 0$, we have
\begin{align*}
    \norm{(\mu_0 - \widehat{\chi}_s(z)) f} \geq \norm{f}^{-1} |\mathrm{Re}\inner{f, \mu_0 - \widehat{\chi}_s(z)f}| \geq \mu_0 \norm{f}
\end{align*}
On the other hand, by estimate \eqref{eq:realtoimaginary}, for any $f \in \mathcal{H}$ with $\mathrm{Re} \inner{f, \widehat{\chi}_s(z) f} \geq 0$ and $\norm{f}=1$, we find
\begin{align*}
   \norm{(\mu_0 - \widehat{\chi}_s(z)) f}^2 &\geq  |\inner{f, \bigr(\mu_0 - \widehat{\chi}_s(z)\bigr)f}|^2 = \bigr( \mathrm{Re}\inner{f,\mu_0 - \widehat{\chi}_s(z) f}\bigr)^2 + (\mathrm{Im}\inner{f,\widehat{\chi}_s(z) f})^2\\
   &\geq (\mathrm{Re} \inner{f,\widehat{\chi}_s(z) f})^2(1+g(z)^{-2}) - 2 \mu_0 \mathrm{Re}\inner{f,\widehat{\chi}_s(z) f} + \mu_0^2.
\end{align*}
Thus minimizing the function $\tau \mapsto \tau^2(1+g(z)^{-2}) - 2 \mu_0 \tau +\mu_0^2$ we obtain
\begin{align}
    \norm{(\mu_0 - \widehat{\chi}_s(z)) f} \geq \frac{\mu_0}{\sqrt{1+g(z)^2}} \norm{f} \label{eq:inverselowerbound}
\end{align}
for any such $f$. We thus conclude that \eqref{eq:inverselowerbound} holds for any $f \in \mathcal{H}$ with $\norm{f} =1$  (as $1+g(z)^2 \geq 1$). Therefore, $\mu_0 - \widehat{\chi}_s(z)$ is injective and the range is closed whenever $g(z) < +\infty$, which is precisely the set $\{ z = \omega + i \eta : \eta \neq 0 \mbox{ or } |\omega|<\omega_1\}$. Moreover, since $\widehat{\chi}_s(z)^\ast = \widehat{\chi}_s(\bar{z})$ (see Proposition~\ref{prop:essentialspec}) and $g(z) = g(\bar{z})$, the adjoint $(\mu_0 - \widehat{\chi}_s(z))^\ast = \mu_0 - \widehat{\chi}_s(\bar{z})$ is also injective, which implies that $\mu_0 - \widehat{\chi}_s(z)$ is invertible. Estimate \eqref{eq:inverseblowuprate} now follows from \eqref{eq:inverselowerbound} and the estimate $g(z) = \max\{ 0, (\omega^2-\eta^2-\omega_1^2)/|\omega \eta|\} \leq |\omega|/|\eta|$. 
\end{proof}

\subsubsection{Inverse of $1-\widehat{\chi}_s(\omega)$ away from the poles of $\widehat{\chi}_s$.}

We now prove a lemma that will be useful to show that all poles of $\bigr(1-\widehat{\chi}_s\bigr)^{-1}$ are simple.
\begin{lem}[Simple poles at discrete spectrum]\label{lem:simple_poles_normal_op} Let $K: B_\epsilon(z_0) \rightarrow \mathcal{B}(\mathcal{H})$ be a holomorphic family of operators such that $K(z_0)$ is normal, $0$ is an isolated point in the spectrum of $K(z_0)$, and $M = {\dim} \ker K(z_0) < \infty$. Suppose that there is a constant $C>0$ such that
\begin{align}
    \norm{K(z_0 + i \eta) f} \geq C|\eta| \norm{f}, \quad  \mbox{ for any $f \in \mathcal{H}$ and $\eta >0 $ close to $0$ .} \label{eq:lowerboundest}
\end{align}
Then $K(z)$ is invertible for $z \neq z_0$ close enough to $z_0$ and
\begin{align*}
    K(z)^{-1} = \frac{K_{-1}}{z-z_0}  + K_0(z),
\end{align*}
where $ \rank K_{-1} = M$ and $K_0(z)$ is holomorphic in $B_\epsilon(z_0)$ (for some possibly smaller $\epsilon>0$).
\end{lem}

\begin{proof} 
First, since $0 \in \sigma_d\bigr(K(z_0)\bigr)$, we know from standard perturbation theory (see Lemma~\ref{lem:continuityofspectra} in the appendix) that for any $\delta>0$ small enough, the projection
\begin{align*}
    Q(z) = \frac{1}{2\pi i} \oint_{\partial B_\delta(0)} (\xi - K(z))^{-1} \mathrm{d} \xi
\end{align*}
is holomorphic for $z$ close enough to $z_0$. Moreover, as $K(z_0)$ is normal, the projection $Q(z_0)$ is the orthogonal projection on $\ker K(z_0)$. Hence, 
\begin{align*}
    \lim_{z\ra z_0} Q(z) K(z) Q(z) \ra P_{\ker K(z_0)} K(z_0) P_{\ker K(z_0)} = 0,
\end{align*}
and therefore, 
\begin{align}
    Q(z) K(z) Q(z) = (z-z_0) K_1 + \mathcal{O}(|z-z_0|^2). \label{eq:Qexp}
\end{align}
 Hence, from \eqref{eq:lowerboundest} and the fact that $Q(z)$ commutes with $K(z)$, we have
\begin{align} 
    \norm{K_1 v} \gtrsim \norm{v}, \quad \mbox{for any $v\in \ran Q(z_0+i \eta)$ and $\eta >0$ small enough.} \label{eq:K1lowerbound}
\end{align}
This implies that $\rank K_1 \geq \rank Q(z) = \rank Q(z_0) = M$. Moreover, since 
\begin{align*}
    K_1 &= \lim_{z \ra z_0}  \frac{1}{z-z_0} Q(z) K(z) Q(z) \\
     &= \lim_{z \ra z_0} Q(z) \Big(\frac{1}{z-z_0} Q(z) K(z) Q(z)\Big) Q(z)\\ 
     &= Q(z_0) K_1 Q(z_0),
\end{align*}
we conclude that $\rank K_1 = \rank Q(z_0) = M< \infty$. But since $\ran Q(z_0)$ is finite dimensional and $Q(z_0)$ commutes with $K_1$, we see that $K_1$ is invertible with respect to $Q(z_0)$. This in turn implies that $Q(z) K(z)Q(z)$ is invertible with respect to $Q(z)$, for $z$ close to $z_0$ excluding $z=z_0$. Indeed, this follows from \eqref{eq:Qexp}, the fact that $Q(z)$ commutes with $Q(z)K(z) Q(z)$, and $\rank Q(z) K(z) Q(z) = \dim \ran Q(z) < \infty$. Hence, we have the decomposition
\begin{align*}
    K(z)^{-1} &= \bigr(Q(z) K(z) Q(z)\bigr)^{-1} + \bigr(\widetilde{Q}(z) K(z) \widetilde{Q}(z)\bigr)^{-1},
\end{align*}
where $\widetilde{Q}(z) = 1 - Q(z)$, and $\bigr(Q(z) K(z) Q(z)\bigr)^{-1}$ respectively $\bigr(\widetilde{Q}(z) K(z) \widetilde{Q}(z)\bigr)^{-1}$ are the inverses with respect to $Q(z)$ and $\widetilde{Q}(z)$. Moreover, by the definition of $Q(z)$, we see that $0 \not \in \sigma(K(z) \bigr\rvert_{\ker {\widetilde{Q}}(z)})$ for any $z$ close to $z_0$. Hence, the inverse $\bigr(\widetilde{Q}(z) K(z) \widetilde{Q}(z)\bigr)^{-1}$ exists and is uniformly bounded (by continuity) around $z=z_0$. We are thus left with computing the pole of $\bigr(Q(z)K(z)Q(z)\bigr)^{-1}$.

To compute the pole of $\bigr(Q(z)K(z)Q(z)\bigr)^{-1}$, first note that from the expansion \eqref{eq:Qexp} and the bound \eqref{eq:K1lowerbound}, we find that $\norm{\bigr(Q(z) K(z) Q(z)\bigr)^{-1}}\lesssim |z-z_0|^{-1}$. Thus, if we multiply \eqref{eq:Qexp} by $\bigr(Q(z) K(z) Q(z)\bigr)^{-1}$ on the left and by $K_1^{-1}$ on the right, we obtain
\begin{align*}
    Q(z) K_1^{-1} = (z-z_0) \bigr(Q(z) K(z) Q(z)\bigr)^{-1} Q(z_0) + \mathcal{O}(|z-z_0|).
\end{align*}
Hence, using that $Q(z) = Q(z_0) + \mathcal{O}(z-z_0)$ (which holds since $Q(z)$ is holomorphic), we conclude that
\begin{align}
    \bigr(Q(z) K(z) Q(z)\bigr)^{-1} = (z-z_0)^{-1} K_1^{-1} + \mathcal{O}(1),
\end{align}
which completes the proof. Note that the remainder $\mathcal{O}(1)$ is holomorphic since the inverse of an holomorphic operator-valued function (whenever defined) is also holomorphic and any holomorphic operator which is uniformly bounded around $z_0$ can be extended to $z_0$ (by Cauchy's formula). 
\end{proof}

\subsubsection{Inverse of $1-\widehat{\chi}_s(\omega)$ at the poles of $\widehat{\chi}_s$.}

We now come to the last difficulty of the proof, namely, dealing with the points $\omega = \omega_j$. The key idea here is to use the operator
\begin{align}
    \widehat{\chi}_{s,j}(z) \coloneqq P_{V_j}^\perp \widehat{\chi}_{s}(z)P_{V_j}^\perp \in \mathcal{B}(V_j^\perp,V_j^\perp) \label{eq:Ljdef}
\end{align}
as a reference for separating the spectrum of $\widehat{\chi}_s$. Precisely, we show here that $\mu_0>0$ is an eigenvalue of $\widehat{\chi}_{s,j}(\omega_j)$ with multiplicity $M$ if and only if the spectrum of $\widehat{\chi}_s(z)$ close to $\mu_0$ converges to $\{\mu_0\}$ as $z \ra \omega_j$ and the associated Riesz projection has rank $M$. Before we show this however, we need one technical lemma. This lemma provides an asymptotic expansion for an continuous operator-valued function close to one of its poles.

\begin{lem}[Inverse of operator-valued function around a pole]\label{lem:inversewithpole} 
    Let $V \subset \mathcal{H}$ be a closed subspace, and let $B\in \mathcal{B}(\mathcal{H})$ be invertible with respect to the orthogonal projection $P_V$. 
    Let $z \mapsto A(z) \in \mathcal{B}(\mathcal{H})$ be an analytic family of operators and suppose that $P_V^\perp A(z) P_V^\perp$ is invertible with respect to $P_V^\perp$, with a uniform bound. 
    Then, for $z \in \C$ small enough, the operator $A(z) + z^{-1} B$ is invertible and we have
\begin{align}
    (A(z) + z^{-1} B)^{-1} = (P_V^\perp A(z) P_V^\perp)^{-1} + \cO(|z|). \label{eq:inversewithpole}
\end{align}
%where the remainder depends on the norms $\norm{G}, \norm{B^{-1}}$ and $\norm{A}$.
\end{lem}

\begin{proof}
    The proof relies on a Schur complement. Writing the operator $A(z) + z^{-1} B$ by blocks, we have 
    \[
    \begin{pmatrix}
        z^{-1}B + P_V A(z) P_V & P_V A(z) P_V^\perp \\
        P_V^\perp A(z) P_V & P_V^\perp A(z) P_V^\perp
    \end{pmatrix} =: \begin{pmatrix}
        \mathcal{A} & \mathcal{B} \\
        \mathcal{C} & \mathcal{D}
    \end{pmatrix}.
    \]
    The Schur complement is the operator $\mathcal{A} - \mathcal{B}\mathcal{D}^{-1}\mathcal{C}$. Using that $z \mathcal{A}$ and $\mathcal{B}\mathcal{D}^{-1}\mathcal{C}$ are uniformly bounded, the Schur complement is invertible for $|z|$ sufficiently small:
    \begin{align*}
        (\mathcal{A} - \mathcal{B}\mathcal{D}^{-1}\mathcal{C})^{-1} &= \mathcal{A}^{-1} \sum_{k \geq 0} \big( z\mathcal{B}\mathcal{D}^{-1}\mathcal{C} (z\mathcal{A})^{-1} \big).
    \end{align*}
    The result then follows from the formula of the inverse of $A(z)+z^{-1}B$ in terms of the blocks
    \begin{align*}
       (A(z)+z^{-1}B)^{-1} &= \begin{pmatrix}
            (\mathcal{A} - \mathcal{B}\mathcal{D}^{-1}\mathcal{C})^{-1} & - (\mathcal{A} - \mathcal{B}\mathcal{D}^{-1}\mathcal{C})^{-1} \mathcal{B} \mathcal{D}^{-1} \\
            - \mathcal{D}^{-1} \mathcal{C}(\mathcal{A} - \mathcal{B}\mathcal{D}^{-1}\mathcal{C})^{-1} & \mathcal{D}^{-1} +\mathcal{D}^{-1} \mathcal{C}(\mathcal{A} - \mathcal{B}\mathcal{D}^{-1}\mathcal{C})^{-1} \mathcal{B} \mathcal{D}^{-1}
       \end{pmatrix}  \\
       &= \begin{pmatrix}
        0 & 0 \\
        0 & (P_V^\perp A(z) P_V^\perp)^{-1},
       \end{pmatrix}+\cO(|z|).
    \end{align*}
\end{proof}

We can now prove the previously mentioned correspondence between the spectra of $\widehat{\chi}_s(z)$ and $\widehat{\chi}_{s,j}(\omega_j)$ as $z$ approaches $\omega_j$. 
\begin{lem}[Convergence of discrete spectra]\label{lem:spectraatexcitation} Let $\widehat{\chi}_{s,j}(z)$ be the operator defined in \eqref{eq:Ljdef}. Then, for any $\mu_0>0$ and $\delta>0$ small enough there exists a neighborhood $U_\delta$ of $\omega_j$ such that $\partial B_\delta(\mu_0) \cap \sigma\bigr(\widehat{\chi}_s(z)\bigr) = \emptyset$ for any $z \in U_\delta$ and 
\begin{align}
	&Q(z) = \frac{1}{2\pi i} \oint_{\partial B_\delta(\mu_0)} (\mu - \widehat{\chi}_s(z))^{-1} \mathrm{d} \mu  = \frac{1}{2\pi i} \oint_{\partial B_\delta(\mu_0)} \bigr(\mu P_{V_j}^\perp - \widehat{\chi}_{s,j}(z)\bigr)^{-1} \mathrm{d} \mu + \mathcal{O}(|z-\omega_j|),
\end{align}
where $\bigr(\mu P_{V_j}^\perp - \widehat{\chi}_{s,j}(z)\bigr)^{-1}$ is the inverse with respect to $P_{V_j}^\perp$. In particular, 
\begin{align*}
    \rank Q(z) = \dim \ker (\mu - \widehat{\chi}_{s,j}(\omega_j)), \quad \mbox{for any $z \in U_\delta$}
\end{align*}
and
\begin{align}
    &Q(z) P_{V_j} = \mathcal{O}(|z-\omega_j|) \quad\mbox{and}\quad P_{V_j} Q(z) = \mathcal{O}(|z-\omega_j|) \label{eq:projexpV_j}
\end{align}
for any $z \in U_\delta$.
\end{lem}
\begin{proof} The first step is to prove the following claim: for any $\mu_0 \in \C$ in the resolvent set of $\widehat{\chi}_{s,j}(\omega_j)$ (where the resolvent/spectra is with respect to $\mathcal{B}(V_j^\perp)$), there exist neighbourhoods $U$ of $\omega_j$ and $W$ of $\mu_0$ such that $W \cap \sigma\bigr(\widehat{\chi}_s(z)\bigr) = \emptyset$ for any $ z\in U$.

So let $\mu_0 \in \C$ belong to the resolvent set of $\widehat{\chi}_{s,j}(\omega_j)$. Then, from standard perturbation theory (see Lemma~\ref{lem:continuityofspectra}) and the continuity of $z \mapsto \widehat{\chi}_{s,j}(z)$, we can find $\delta>0$ small enough such that $B_\delta(\mu_0) \subset \C$ lies on the resolvent set of $\widehat{\chi}_{s,j}(z)$ for any $z$ close enough to $\omega_j$. In particular, the inverse $\bigr(\mu P_{V_j}^\perp - \widehat{\chi}_{s,j}(z)\bigr)^{-1}$ exists and is uniformly bounded for $z$ close enough to $\omega_j$ and $\mu \in W \coloneqq B_{\delta/2}(\mu_0)$. Therefore, we can apply Lemma~\ref{lem:inversewithpole} to $ \mu - \widehat{\chi}_s(z)  = A(z,\mu) + (z-\omega_j)^{-1} B$, where 
\begin{align*}
	A(z,\mu) = \mu - \widehat{\chi}_s(z) + (z-\omega_j)^{-1} B_j, \quad B = B_j,\quad\mbox{and}\quad P_{V_j}^\perp A(z,\mu) P_{V_j}^\perp = \mu P_{V_j}^\perp - \widehat{\chi}_{s,j}(z),
\end{align*}
 to conclude that $\mu - \widehat{\chi}_s(z)$ is invertible and 
\begin{align}
 	\bigr(\mu - \widehat{\chi}_s(z)\bigr)^{-1} = \bigr(\mu P_{V_j}^\perp - \widehat{\chi}_{s,j}(z)\bigr)^{-1} + \mathcal{O}(|z-\omega_j|), \label{eq:projexp1}
, \quad\mbox{for any $\mu \in W$ and $z \in U$,}\end{align}
 where $(\mu P_{V_j}^\perp - \widehat{\chi}_{s,j})^{-1}$ is the inverse with respect to $P_{V_j}^\perp$. 
 %Note that the operator $A= A(z,\mu)$ here depends on $\mu$ and $z$, which is not the case in Lemma~\ref{lem:inversewithpole}. However, both $A(z,\mu)$ and $(\mu P_{V_j}^\perp - \widehat{\chi}_{s,j}(z))^{-1}$ are uniformly bounded for $z$ close to $\omega_j$ and $\mu \in W$, and therefore, the constant $C>0$ on the remainder of \eqref{eq:projexp1} can be taken independent of $z$ and $\mu$. We have thus proved our claim.
 
 Next, let {$0 < \mu_0 \in \sigma\bigr(\widehat{\chi}_{s,j}(\omega_j)\bigr)$}. So, by the same arguments in the proof of Proposition~\ref{prop:essentialspec}, we find that $\mu_0$ belongs to the discrete spectrum of $\widehat{\chi}_{s,j}(\omega_j)$. Thus, from classical perturbation theory again, we can find an annulus 
\begin{align*}
	\overline{B_{\delta}(\mu_0)\setminus B_{\delta/2}(\mu_0)} \subset \{ z \in \C : \mathrm{Re}(z) >0 \} 
\end{align*}
contained in the resolvent set of $\widehat{\chi}_{s,j}(z)$ for any $z$ sufficiently close to $\omega_j$. In particular, from our first claim and a compactness argument, we can take a small enough neighbourhood of $z=\omega_j$ such that the annulus $\overline{B_{\delta}(\mu_0)\setminus B_{\delta/2}(\mu_0)}$ also lies inside the resolvent set of $\widehat{\chi}_s(z)$. In this neighbourhood, the spectral (Riesz) projection defined by
\begin{align*}
	Q(z) = \frac{1}{2\pi i} \oint_{\partial B_{\delta}(\mu_0)} \bigr(\mu - \widehat{\chi}_s(z)\bigr)^{-1} \mathrm{d} \mu
\end{align*}
is holomorphic and has constant rank. Moreover, substituting $\bigr(\mu- \widehat{\chi}_s(z)\bigr)^{-1}$ by \eqref{eq:projexp1} in the above we obtain
\begin{align}
    Q(z) = \frac{1}{2\pi i} \oint_{\partial B_\delta(\mu_0)} \bigr(\mu P_{V_j}^\perp - \widehat{\chi}_{s,j}(z)\bigr)^{-1} \mathrm{d} \mu + \mathcal{O}(|z-\omega_j|). \label{eq:projexp}
\end{align}
Moreover, since $\bigr(\mu P_{V_j}^\perp - \widehat{\chi}_{s,j}(z)\bigr)^{-1} P_{V_j} = P_{V_j} \bigr(\mu P_{V_j}^\perp - \widehat{\chi}_{s,j}(z)\bigr)^{-1} = 0$ we obtain \eqref{eq:projexpV_j}. 

To complete the proof we just need to show that $\rank Q(z) = \dim \ker \mu_0 - \widehat{\chi}_{s,j}(\omega_j)$. This follows from the fact that $Q(z)$ is a continuous family of projections, hence $\rank Q(z)$ is constant, and $Q(z) \ra \frac{1}{2\pi i} \oint \bigr(\mu P_{V_j}^\perp - \widehat{\chi}_{s,j}(\omega_j)\bigr)^{-1}$, which is the orthogonal projection on $\ker \mu_0 - \widehat{\chi}_{s,j}(\omega_j)$ (as $\widehat{\chi}_{s,j}(\omega_j)$ is symmetric). 
\end{proof}

We are now in position to prove Proposition~\ref{prop:1-chi_s_meromorphic}.
    \begin{proof}[Proof of Proposition~\ref{prop:1-chi_s_meromorphic}]
        Note that the set $\{ z \in \mathcal{D}_\Omega \setminus \mathcal{P}(\widehat{\chi}_s): 1 \not \in \sigma(\widehat{\chi}(z) \} \subset \C$ is open by continuity. Hence, $(1-\widehat{\chi}_s(z)\bigr)^{-1}$ is well-defined and holomorphic on this set, and we just need to worry about the points where $1-\widehat{\chi}_s$ is not invertible, and the points $\omega_j$ where $\chi_s$ blows-up. By Lemma~\ref{lem:inverseawayofimaginary}, the set of points where $1-\widehat{\chi}_s$ is not invertible is contained in the intervals $\omega \in (-\Omega, -\omega_1] \cap [\omega_1,\Omega)$. Moreover, by the symmetries of $\widehat{\chi}_s$, it is enough to look on the positive interval $[\omega_1,\Omega)$.
        
        So first, let us consider the points $\omega \in (\omega_j, \omega_{j+1})$ for some $j\leq m$, where $1-\widehat{\chi}_s(\omega)$ is not invertible. For these points, we know from Proposition~\ref{prop:essentialspec} that $1$ belongs to the discrete spectrum of $\widehat{\chi}_s(\omega)$. Hence, from Lemma~\ref{lem:simple_poles_normal_op}, estimate~\eqref{eq:inverseblowuprate}, and the fact that $\widehat{\chi}_s(\omega)$ is self-adjoint, we conclude that this set of points is discrete, and that $\bigr(1-\widehat{\chi}_s(\omega)\bigr)^{-1}$ has a pole with rank equals to $\dim \ker \bigr(1-\widehat{\chi}_s(\omega)\bigr)$ at any such point.
        
        Next, we want to show that any excitation $\omega_j$ is a pole of $\bigr(1-\widehat{\chi}_s\bigr)^{-1}$ with rank equals to $\dim \ker \bigr(1- P_{V_j}^\perp \widehat{\chi}_s(\omega_j) P_{V_j}^\perp\bigr)$. So let $\omega_j$ be a pole of $\widehat{\chi}_s$, and let $z$ be in a neighborhood of $\omega_j$ such that the projection 
        \begin{equation}
            Q(z) = \frac{1}{2\pi i} \oint_{\partial B_\delta(1)} (\xi - \widehat{\chi}_s(z))^{-1} \mathrm{d}\xi,
        \end{equation}
        has rank $M_j = \dim \ker \bigr(1-P_{V_j}^\perp\widehat{\chi}_s(\omega_j)P_{V_j}^\perp\bigr)$. That this projection is well-defined and holomorphic for $z$ close to $\omega_j$ is a consequence of Lemma~\ref{lem:spectraatexcitation}. Thus, since $\widehat{\chi}_s(z)$ commutes with $Q(z)$, we can deal with the operators 
\begin{align*}
    1-\widehat{\chi}_s(z) = \underbrace{Q(z)\bigr(1-\widehat{\chi}_s(z)\bigr) Q(z)}_{\coloneqq K(z)} + \underbrace{\widetilde{Q}(z) \bigr(1-\widehat{\chi}_s(z)\bigr) \widetilde{Q}(z)}_{\coloneqq \widetilde{K}(z)}
\end{align*}
separately. 

Let us start with $K(z)$. From the definition of $Q(z)$ and Lemma~\ref{lem:spectraatexcitation}, we see that $K(z) \ra 0$ as $z \ra \omega_j$. Hence, $K(z)$ can be expanded as
\begin{align*}
    K(z) = (z-\omega_j) K_1 + \mathcal{O}(|z-\omega_j|),
\end{align*}
for some $K_1 \in \mathcal{B}(\mathcal{H})$. Moreover, from the blow-up estimate \eqref{eq:inverseblowuprate}, we also have
\begin{align*}
    \norm{K(\omega_j + i \eta ) f} \gtrsim |\eta| \norm{f}, \quad \mbox{ for any $f \in \ran Q(z)$ and $\eta$ small.}
\end{align*}
Hence, the same arguments from the proof of Lemma~\ref{lem:simple_poles_normal_op} leads to the conclusion that $K_1$ is invertible with respect to $Q(\omega_j)$, that $K(z)$ is invertible with respect to $Q(z)$ for $z\neq \omega_j$, and that
\begin{align}
    K(z)^{-1} = (z-\omega_j)^{-1} K_{-1} + \mathcal{O}(1), \label{eq:Kinverse}
\end{align}
for some operator $K_{-1}$ with $\rank K_{-1} = M = \dim \ker (1- P_{V_j}^\perp \widehat{\chi}_s(\omega_j) P_{V_j}^\perp)$. The big-O term here is with respect to the limit $z\ra \omega_j$.

For $\widetilde{K}(z)$ we can use formula \eqref{eq:inverse1-Pformula} in the appendix. Indeed, introducing the Riesz projection $Q_j(z) = \oint_{\partial B_\delta(\mu_0)} \bigr(\mu P_{V_j}^\perp - \widehat{\chi}_{s,j}(z)\bigr)^{-1} \mathrm{d} \mu$ of $\widehat{\chi}_{s,j}(z)$ around $1$ 
then from the definition of $\widetilde{Q}(z)$, formula \eqref{eq:inverse1-Pformula}, and Lemma~\ref{lem:inversewithpole}, we find that
\begin{align}
    \widetilde{K}(z)^{-1} &= \frac{1}{2\pi i} \oint_{\partial B_\delta(1)} \frac{1}{\mu-1} \bigr(\mu - \widehat{\chi}_s(z)\bigr)^{-1} \mathrm{d} \mu \nonumber \\
    &= \frac{1}{2 \pi i } \oint_{\partial B_\delta(1)} \frac{1}{\mu-1} \bigr(\mu P_{V_j}^\perp - \widehat{\chi}_{s,j}(z)\bigr)^{-1} \mathrm{d} \mu + \mathcal{O}(|z-\omega_j|) \nonumber \\
    & = \biggr(\bigr(P_{V_j}^\perp- Q_j(z)\bigr)\bigr(1-\widehat{\chi}_{s,j}(z)\bigr)\bigr(P_{V_j}^\perp - Q_j(z)\bigr)\biggr)^{-1} + \mathcal{O}(|z-\omega_j|) = \mathcal{O}(1). \label{eq:Ktildeinverse}
\end{align}
Combining \eqref{eq:Kinverse} and \eqref{eq:Ktildeinverse}, we have shown that 
\begin{align*}
    (1-\widehat{\chi}_s(z)\bigr)^{-1} =  K^{-1} + \widetilde{K}^{-1} =\frac{K_{-1}}{z-\omega_j} + K_0(z),
\end{align*}
where $K_{-1}$ is invertible with respect to the orthogonal projection on $\ker 1-\widehat{\chi}_{s,j}(\omega_j)$ and the operator $K_0(z)$ is holomorphic and uniformly bounded around $z=\omega_j$.

To complete the proof, it is enough to show that $\ran K_0(\omega_j) \subset \ran K_{-1}$. This follows from the identity
\begin{align*}
    K_0(\omega_j)P_{V_j} &= \lim_{z\ra \omega_j} \bigr(1-\widehat{\chi}_s(z)\bigr)^{-1} P_{V_j} = \lim_{z \ra \omega_j} K(z)^{-1} P_{V_j} + \underbrace{\lim_{z \ra \omega_j} \bigr(\widetilde{K}(z)\bigr)^{-1} P_{V_j}}_{=0} \\
    &= \lim_{z\ra \omega_j} Q(z) K(z)^{-1} P_{V_j} = Q(\omega_j) \lim_{z \ra \omega_j} K(z)^{-1} P_{V_j},
\end{align*}
where we used that $K_{-1} P_{V_j} =0$, and $\widetilde{K}(z)^{-1} P_{V_j} = \mathcal{O}(|z-\omega_j|)$ (see \eqref{eq:Ktildeinverse}), and the fact that $\ran Q(\omega_j) = \ran K_{-1}$.
\end{proof}

\subsection{Proof of Proposition~\ref{prop:forwardshiftchis}}

The strategy here is to show that the eigenvalues of $\widehat{\chi}_s(\omega)$, as function of $\omega$, are strictly decreasing along the intervals $(\omega_j,\omega_{j+1})$, and then analyze what happens when $\omega$ crosses  $\omega_j$. 
For this, let us introduce the number of eigenvalues, counting multiplicity, greater than $\mu_0$ by
\begin{align}
    n_{\mu_0}(\omega) \coloneqq \sum_{\mu > \mu_0} \dim \ker \bigr(\mu - \widehat{\chi}_s(\omega)\bigr) \label{eq:ndef}.
\end{align}
Then, the following lemma holds.

\begin{lem}[Strictly decreasing eigenvalues]  \label{lem:strictlydecreasing} The positive eigenvalues of $\widehat{\chi}_s(\omega)$ are decreasing functions of $\omega$ in the interval $(\omega_j,\omega_{j+1})$, $j \geq 0$. Moreover, for any $\mu_0 > 0$, we have
\begin{align}
    n_{\mu_0}(\omega_j^+) =  n_{\mu_0}(\omega_j^-) + \dim V_j - \dim \ker \bigr(\mu_0-P_{V_j}^\perp \widehat{\chi}_s(\omega_j) P_{V_j}^\perp \bigr) , \label{eq:crossingest}
\end{align}
where $n_{\mu_0}(\omega_j^+) = \lim_{\omega \ra \omega_j,\omega>\omega_j} n_{\mu_0}(\omega)$ and $n_{\mu_0}(\omega_j^-) = \lim_{\omega \ra \omega_j,\omega<\omega_j} n_{\mu_0}(\omega)$ are respectively the right and left limits of $n_{\mu_0}(\omega)$ at $\omega_j$.
\end{lem}
\begin{proof} 
    We first note that since the function {$h(\lambda,\omega)  = \frac{2\lambda}{\omega^2-\lambda^2}$} is  decreasing for $\omega$ in the intervals $(-\infty,\lambda)$ and $(\lambda,\infty)$, and since the spectral measure $\norm{P_{\lambda+E_0} T^\ast f}^2$ is not identically zero for any $f$ {in the complement of the kernel of $T^*$}, then for any $\omega > \omega'$ in $(\omega_j,\omega_{j+1})$
\begin{align}
   \inner{f, \widehat{\chi}_s(\omega') f} = \int_{\omega_1}^\infty h(\lambda, \omega') \mathrm{d} \norm{P^{H}_{\lambda+E_0} T^\ast f}^2 > \int_{\omega_1}^\infty h(\lambda,\omega) \mathrm{d} \norm{P^{H}_{\lambda+E_0} T^\ast f}^2 = \inner{f,\widehat{\chi}_s(\omega) f}. \label{eq:strictlydecreasing}
\end{align}
Hence by Rayleigh-Ritz principle, the positive eigenvalues of $\widehat{\chi}_s(\omega)$ are decreasing functions of $\omega$ in the interval $(\omega_j,\omega_{j+1})$.
%Next, note that the positive eigenvalues of $\mu_1(\omega) \geq \mu_2(\omega) \geq ... > 0$ of $\widehat{\chi}_s(\omega)$ are given by the max-min principle
%\begin{align}
%    \mu_k(\omega) = \sup_{\substack{W\subset \mathcal{H} \\ \dim W = k}} \underline{\mu}(\omega,W), . \label{eq:maxminprinciple}
%\end{align}
%where $\underline{\mu}(\omega,W) \coloneqq \inf \{ \inner{f,\widehat{\chi}_s(\omega) f} : f \in W, \norm{f} =1 \}$. Then since for any finite-dimensional $W\subset \mathcal{H}$ the inf in $\underline{\mu}(\omega,W)$ is attained (by compactness) at some $f_W$, we find that
%\begin{align}
%    \underline{\mu}(\omega',W) = \inner{f_W,\widehat{\chi}_s(\omega)f_W} > \inner{f_W, \widehat{\chi}_s(\omega) f_W} \geq \underline{\mu}(\omega,W), \label{eq:intermediatemuest}
%\end{align}
%for any $\omega > \omega'$ in $(\omega_j,\omega_{j+1})$. Similarly, we know that the sup in \eqref{eq:maxminprinciple} is also attained for some $W_k \subset \mathcal{H}$. Thus we conclude from \eqref{eq:intermediatemuest} that 
%\begin{align*}
%    \mu_k(\omega) = \underline{\mu}(\omega,W_k) < \underline{\mu}(\omega',W_k) \leq \mu_k(\omega'),
%\end{align*}
%for any $\omega' > \omega$ in $(\omega_j,\omega_{j+1})$, and therefore, the eigenvalues are strictly decreasing. 

The existence of the right and left limits of $n_{\mu_0}(\omega)$ at $\omega_j$ follows since $n_{\mu_0}(\omega)$ is decreasing. 
By Lemma~\ref{lem:spectraatexcitation} and the decreasing property of the eigenvalues of $\widehat{\chi}_s$, the left limit of $n_{\mu_0}(\omega_j^-)$, $\omega<\omega_j$ is exactly the number of eigenvalues of $P_{V_j}^\perp \widehat{\chi}_s(\omega_j) P_{V_j}^\perp$ equal to or greater than $\mu_0$.
Indeed, by Lemma~\ref{lem:spectraatexcitation}, an eigenvalue of $P_{V_j}^\perp \widehat{\chi}_s(\omega_j) P_{V_j}^\perp$ equal to or greater than $\mu_0$ corresponds to another eigenvalue of $\widehat{\chi}_s(\omega)$, for $\omega$ in a neighborhood of $\omega_j$. Since $\omega < \omega_j$, by the decreasing property, the corresponding eigenvalue is greater than $\mu_0$. 
Conversely, if $\mu(\omega)$ is an eigenvalue of $\widehat{\chi}_s(\omega)$ such that $\lim_{\omega \to \omega_j, \omega < \omega_j} \mu(\omega) \geq \mu_0$, then as $\lim_{\omega \to \omega_j, \omega < \omega_j} \frac{2\omega_j}{\omega^2-\omega_j^2} = -\infty$ the corresponding family of eigenfunctions has a vanishing component on $V_j$. Hence $\mu(\omega)$ converges to an eigenvalue of $P_{V_j}^\perp \widehat{\chi}_s(\omega_j) P_{V_j}^\perp$. 

For the right limit $n_{\mu_0}(\omega_j^+)$, consider an eigenvalue $\mu(\omega)$ of $\widehat{\chi}_s(\omega)$, $\omega > \omega_j$ greater than $\mu_0$. Since $\omega \mapsto \mu(\omega)$ is decreasing, it either diverges at $\omega_j$ or it converges to an eigenvalue of $P_{V_j}^\perp \widehat{\chi}_s(\omega_j) P_{V_j}^\perp$ strictly greater than $\mu_0$ by a similar argument as above. 
Finally there are exactly $\dim V_j$ eigenvalues blowing up at $\omega_j$ since for any $f \in V_j$, we have 
\begin{align*}
    \inner{f,\widehat{\chi}_{s}(\omega) f} \geq\inner{f,\frac{2\omega_j}{\omega^2-\omega_j^2}B_j f} - \norm{\widehat{\chi}_s(\omega) - \frac{2\omega_j}{\omega^2-\omega_j^2} B_j}\norm{f} \gtrsim \frac{2\omega_j}{\omega^2-\omega_j^2} \norm{f}^2  ,
\end{align*}
so $\lim_{\omega \to \omega_j, \omega > \omega_j} \inner{f,\widehat{\chi}_{s}(\omega) f} = \infty$ and $P_{V_j}^\perp \widehat{\chi}_s(\omega) P_{V_j}^\perp$ is bounded in a neighborhood of $\omega_j$.

\end{proof}

\begin{proof}[Proof of Proposition~\ref{prop:forwardshiftchis}]
From Proposition~\ref{prop:1-chi_s_meromorphic}, $1-\widehat{\chi}_s(\omega)$ is invertible for $0<\omega < \omega_1$.
Since $P_{V_1}^\perp \widehat{\chi}_s(\omega) P_{V_1}^\perp$ is negative for $|\omega| < \omega_2$, from Lemma~\ref{lem:spectraatexcitation}, $\omega_1$ is not a pole of $(1-\widehat{\chi}_s)^{-1}$.
By Proposition~\ref{prop:poleschi0}, we have $\rank_{\omega_j}(\widehat{\chi}_s) = \dim V_j$, so by Lemma~\ref{lem:inverseawayofimaginary} it is enough to show that
\begin{align}
    \sum_{\omega_1 < \omega < \omega_{j+1}} \rank_\omega\bigr((1-\widehat{\chi}_s)^{-1}\bigr) \leq \sum_{k \leq j} \dim V_k \quad \mbox{for any $j \leq m$.} \label{eq:endgoal}
\end{align}
%(Note that $\omega_1$ is not a pole of $(1-\widehat{\chi}_s)^{-1}$ since $P_{V_1}^\perp \widehat{\chi}_s(\omega) P_{V_1}^\perp$ is negative for $|\omega| < \omega_2$.)
By Proposition~\ref{prop:1-chi_s_meromorphic}, $\rank_\omega\bigr((1-\widehat{\chi}_s)^{-1}\bigr) = \dim \ker \bigr(1-\widehat{\chi}_s(\omega)\bigr)$ for any $\omega \in (\omega_j,\omega_{j+1})$. From the decreasing property of the eigenvalues in Lemma~\ref{lem:strictlydecreasing}, the sum of the ranks of the poles in the interval $(\omega_j,\omega_{j+1})$ is given by the number of eigenvalues of $\widehat{\chi}_s(\omega)$ that cross 1. Hence for any $j \geq 0$,
\begin{align*}
    \sum_{\omega_j < \omega < \omega_{j+1}} \rank_\omega\bigr((1-\widehat{\chi}_s)^{-1}\bigr) = n_{1}(\omega_j^+) - n_1(\omega_{j+1}^-).
\end{align*}
As a result, combining Lemma~\ref{lem:spectraatexcitation}, the estimate~\eqref{eq:crossingest} and the rank charaterization in~\eqref{eq:rankpoleschis} we get
\begin{align*}
    \sum_{\omega_1<\omega < \omega_{j+1}} \rank_\omega\bigr((1-\widehat{\chi}_s)^{-1}\bigr) &= \sum_{k=1}^j \biggr( \rank_{\omega_k}\bigr((1-\widehat{\chi}_s)^{-1}\bigr) + \sum_{\omega_k < \omega < \omega_{k+1}} \rank_\omega\bigr((1-\widehat{\chi}_s)^{-1}\bigr)\biggr) \\
    &= \sum_{k=1}^j \biggr( \dim \ker \bigr(\mu_0-P_{V_j}^\perp \widehat{\chi}_s(\omega_j) P_{V_j}^\perp \bigr) +  n_{1}(\omega_k^+) - n_1(\omega_{k+1}^-)\biggr) \\
    & = \sum_{k=1}^j n_1(\omega_k^-) + \dim V_k  - n_1(\omega_{k+1}^-) \\
    &= n_1(\omega_1^-) - n_1(\omega_{j+1}^-) + \sum_{k=1}^j \dim V_k .
\end{align*}
But since $n_1(\omega_{j+1}^+) \geq 0$ and $n_1(\omega_1^-) = 0$ as $\widehat{\chi}_s(\omega)$ is nonpositive-semidefinite for $\omega < \omega_1$, we obtain \eqref{eq:endgoal}.
\end{proof}

\section{The Fourier transform of $\chi^\mathrm{RPA}$}
\label{sec:fourier_chi_rpa}

We can now use Propositions~\ref{prop:1-chi_s_meromorphic} and~\ref{prop:forwardshiftchis} {in the particular case $T = F_H^{1/2} S$} to prove Theorems~\ref{theo:polesRPA} and \ref{theo:rankpolesRPA}.

\begin{proof}[Proof of Theorem~\ref{theo:polesRPA}] Let $\chi_0$ be the density-density response function of some Hamiltonian satisfying Assumption~\ref{assump:groundstate} and let $\chi^\mathrm{RPA} = \mathcal{S}^{RPA}(\chi_0)$ be the associated solution to the RPA-Dyson equation. Then since $\norm{\chi_0(t)}_{L^2+L^\infty,L^1\cap L^2}$ is uniformly bounded by Proposition~\ref{prop:Lpregularity}, from the Gronwall inequality we know that $\norm{\chi^\mathrm{RPA}(t)}_{L^2+L^\infty,L^1\cap L^2} \lesssim e^{Dt}$ for some $D>0$. Hence, the Fourier transform $\widehat{\chi^\mathrm{RPA}}(z)$ is well-defined for $\mathrm{Im}(z) > D$ and we have (by the convolution property of the Fourier transform)
\begin{align}
    \widehat{\chi^\mathrm{RPA}}(z) = \widehat{\chi_0}(z) + \widehat{\chi_0}(z) F_H \widehat{\chi^\mathrm{RPA}}(z), \quad \mbox{for $\mathrm{Im}(z) > D$.} \label{eq:Fourierdyson}
\end{align}
Let $\widetilde{\chi}$, $\widetilde{\chi_0}$ and $\widehat{\chi}_s$ be the operators defined by
\begin{align}
    \widetilde{\chi} \coloneqq F_H^{\frac12} \widehat{\chi^\mathrm{RPA}}, \quad  \widetilde{\chi_0} \coloneqq F_H^{\frac12} \widehat{\chi_0}, \quad \mbox{and} \quad  \widehat{\chi}_s \coloneqq F_H^{\frac12} \widehat{\chi}_0 F_H^{\frac12}, \label{eq:operatordef}
\end{align}
where $F_H^{\frac12} = \sqrt{4\pi} (-\Delta)^{-\frac12}$ is up to a multiplicative constant the convolution against $1/|\cdot|^2$. Then we have 
\begin{align*}
    \widetilde{\chi}(z) = \widetilde{\chi}_0(z) + \widehat{\chi}_s(z) \widetilde{\chi}(z) \quad \Rightarrow \quad \bigr(1-\widehat{\chi}_s(z)\bigr) \widetilde{\chi}(z) = \widetilde{\chi_0}(z) .
\end{align*}
Moreover, we see from the Hardy-Littlewood-Sobolev inequality that $\widetilde{\chi_0}(z) \in \mathcal{B}\bigr(L^2(\R^3)+L^\infty(\R^3),L^2(\R^3)\bigr)$, and that $\widehat{\chi}_s$ is an operator of the form of Equation~\eqref{eq:def_symmetrized_chi} with $T = F_H^\frac12 S \in \mathcal{B}(\mathcal{H}_N,L^2(\R^3))$. Therefore, from Proposition~\ref{prop:1-chi_s_meromorphic}, the map
\begin{align*}
    z\mapsto \bigr(1-\widehat{\chi}_s(z)\bigr)^{-1} \widetilde{\chi}_0(z) \in \mathcal{B}\bigr(L^1(\R^3)\cap L^2(\R^3), L^2(\R^3)\bigr),
\end{align*}
is the unique meromorphic extension of $\widetilde{\chi}$ to the domain $\mathcal{D}_\Omega = \{ z \in \C : \mathrm{Im}(z) \neq 0 \mbox{ or } |\mathrm{Re}(z)|<\Omega \}$. Now going back to Equation~\eqref{eq:Fourierdyson}, we see from \eqref{eq:operatordef} that
\begin{align}
    \widehat{\chi^\mathrm{RPA}}(z) = \widehat{\chi_0}(z) + \widehat{\chi_0}(z)F_H^{\frac12} \widetilde{\chi}(z) = \widehat{\chi}_0(z) + \widehat{\chi}_0(z) F_H^{\frac12} \bigr(1-\widehat{\chi}_s(z)\bigr)^{-1} F_H^{\frac12} \widehat{\chi}_0(z). \label{eq:chirparep}
\end{align}
In particular, $\widehat{\chi^\mathrm{RPA}}$ has an unique meromorphic extension as a map from $\mathcal{D}_\Omega$ to $\mathcal{B}(L^2+L^\infty,L^1\cap L^2)$, which proves item~\ref{it:meromorphicpolesRPA} from Theorem~\ref{theo:polesRPA}.

For the other item in Theorem~\ref{theo:polesRPA}, we want to relate the poles of $\widehat{\chi^\mathrm{RPA}}$ to the poles of $(1-\widehat{\chi}_s)^{-1}$. This can be done by observing that, since $F_H^{\frac12} : L^1(\R^3)\cap L^2(\R^3) \rightarrow L^2(\R^3)$ is injective and bounded, a point $z \in \mathcal{D}_\Omega$ is a pole of $\widehat{\chi^\mathrm{RPA}}(z)$ with finite rank if and only if it is a pole of $\widetilde{\chi} = F_H^{\frac12} \widehat{\chi^\mathrm{RPA}}$ with the same rank. Therefore, if we can show that any pole of $\widetilde{\chi}$ is simple and the inequality 
\begin{align}
    \rank_{\omega}(\widetilde{\chi}) \leq \rank_\omega \bigr((1-\widehat{\chi}_s)^{-1}\bigr) \label{eq:ineqpoles}
\end{align}
holds for any $0< \omega< \Omega$ (with the convention that the rank is zero if $\omega$ is not a pole), then Theorem~\ref{theo:polesRPA} follows from the forward shift property in  Proposition~\ref{prop:forwardshiftchis}.

For $\omega \in \mathcal{P}\big( (1-\widehat{\chi}_s)^{-1} \big) \setminus \mathcal{P}(\widehat{\chi_0})$, since $\widehat{\chi}_0$ is holomorphic in a neighborhood of $\omega$, it is clear from \eqref{eq:chirparep} that $\omega$ is at most a simple pole of $\widehat{\chi^\mathrm{RPA}}$ and that inequality~\eqref{eq:ineqpoles} holds.
We are thus left with the points in $\mathcal{P}(\widehat{\chi_0})$. Let $\omega_j \in \mathcal{P}(\widehat{\chi_0})$. Since $F_H^{\frac12}$ is injective and bounded from $L^1(\mathbb{R}^3)\cap L^2(\mathbb{R}^3)$ to $L^2(\mathbb{R}^3)$, we have $\dim \ran F_H^{\frac12} S P^H_{E_0+\omega_j} = \dim \ran S P^H_{E_0+\omega_j}$. 
This implies that the poles of $\widehat{\chi}_s$ are precisely the poles of $\widehat{\chi_0}$ and they have the same rank. Hence using the spectral decomposition of $H$, we have that $\widetilde{\chi}_0(z) + 2\frac{\omega_j}{\omega_j^2-z^2}  F_H^{\frac12} SP^H_{E_0+\omega_j} S^\ast$ is bounded. By the decomposition~\eqref{eq:detailedpoleinfo} of $(1-\widehat{\chi}_s(z))^{-1}$ for $z$ in a neighborhood of $\omega_j$, we obtain 
\begin{align*}
    (z-\omega_j)\widetilde{\chi}(z) &= \biggr(K_{-1}+ (z-\omega_j) K_0\biggr) \biggr(\frac{F_H^{\frac12} SP^H_{E_0+\omega_j} S^\ast}{z-\omega_j} + \widetilde{\chi}_0 - \frac{F_H^{\frac12} SP^H_{E_0+\omega_j} S^\ast}{z-\omega_j} \biggr) + \mathcal{O}(|z-\omega_j|)  \\
    &= K_{-1} \frac{F_H^{\frac12} SP^H_{E_0+\omega_j} S^\ast}{z-\omega_j}+ K_{-1} \biggr(\widetilde{\chi}_0 - \frac{F_H^{\frac12} SP^H_{E_0+\omega_j} S^\ast}{z-\omega_j} \biggr) + K_0 F_H^{\frac12} SP^H_{E_0+\omega_j} S^\ast \\
    & \qquad \qquad+ \mathcal{O}(|z-\omega_j|).
\end{align*}
Recall that $\widehat{\chi}_s$ is given by Equation~\eqref{eq:def_symmetrized_chi} by taking $T = F_H^{\frac12} S$. Hence using the last statement in Proposition~\ref{prop:1-chi_s_meromorphic}, we get that $K_{-1} F_H^{\frac12} SP^H_{E_0+\omega_j} S^\ast=0$. 
This shows that $\omega_j$ is at most a simple pole of $\widetilde{\chi}$. 
Moreover using again the last statement of Proposition~\ref{prop:1-chi_s_meromorphic},  $\ran K_0 F_H^{\frac12} SP^H_{E_0+\omega_j} \subset \ran K_{-1}$. 
As by Equation~\eqref{eq:rankpoleschis} $\dim \ran K_{-1} = \dim {\ker}\bigr(1-P_{V_j}^\perp \widehat{\chi}_s(\omega_j) P_{V_j}^\perp \bigr) = \rank_{\omega_j}\bigr((1-\widehat{\chi}_s)^{-1}\bigr)$, the inequality $\rank_{\omega_j} (\widetilde{\chi}) \leq \rank_{\omega_j}\bigr((1-\widehat{\chi}_s)^{-1} \bigr)$ holds for any $j\leq m$ and the proof is complete.
\end{proof}

Notice that we have proved that $\rank_\omega (\widehat{\chi^\mathrm{RPA}}) {\le}\rank_\omega \bigr((1-\widehat{\chi}_s)^{-1}\bigr)$ for any $0<\omega < \Omega$, which will be useful in the proof of Theorem~\ref{theo:rankpolesRPA}.

\begin{proof}[Proof of Theorem \ref{theo:rankpolesRPA}] By the rank characterization in Equation~\eqref{eq:rankpoleschis} from Proposition~\ref{prop:1-chi_s_meromorphic}, it is enough to show that $\rank_\omega (\widehat{\chi^\mathrm{RPA}}) = \rank_\omega \bigr((1-\widehat{\chi}_s)^{-1}\bigr)$ for any $0<\omega < \Omega$. The inequality $\rank_\omega(\widehat{\chi^\mathrm{RPA}}) = \rank_\omega (\widetilde{\chi})  \leq \rank_{\omega} \bigr((1-\widehat{\chi}_s)\bigr)$ has already been proved . For the converse inequality, observe that, since $\widetilde{\chi}:\mathcal{D}_\Omega \rightarrow \mathcal{B}(L^2+L^\infty,L^2)$ is meromorphic, the map $\widetilde{\chi} F_H^{\frac12} : \mathcal{D}_\Omega \rightarrow \mathcal{B}(L^2,L^2)$ is also a meromorphic family of operators. Thus the inquality $\rank_\omega (\widetilde{\chi} F_H^{\frac12}) \leq \rank_\omega (\widetilde{\chi}) $ holds since the composition of linear operators can only lower the rank. The proof is now complete because
\begin{align}
    \widetilde{\chi}(z) F_H^{\frac12} = \bigr(1-\widehat{\chi}_s(z)\bigr)^{-1} \underbrace{F_H^{\frac12} \widehat{\chi}_0(z) F_H^{\frac12}}_{= \widehat{\chi}_s(z)} = \bigr(1-\widehat{\chi}_s(z)\bigr)^{-1} - 1, \label{eq:niceidentity}
\end{align}
which implies that $ \rank_\omega\bigr((1-\widehat{\chi}_s)^{-1}\bigr) = \rank_\omega(\widetilde{\chi} F_H^{\frac12}) \leq \rank_\omega(\widehat{\chi^\mathrm{RPA}})$.
\end{proof}

\begin{note} Ideally, we would like to conclude the proof of Theorem~\ref{theo:polesRPA} directly from identity~\eqref{eq:niceidentity}. However, this is not possible because $F_H^{\frac12} : L^2 \rightarrow L^2 + L^\infty$ is not surjective (nor is the image dense in $L^2 + L^\infty$), and some poles could in principle vanish when composing $\widehat{\chi^\mathrm{RPA}}$ with $F_H^{\frac12}$ on the right.\end{note}

\appendix

\section{Time-dependent density functional theory}
\label{sec:tddft_short_intro}

In TDDFT, one postulates the existence of an (exact) time-dependent exchange-correlation potential\footnote{For proofs of the existence respectively uniqueness up to a time-dependent constant of such a potential see \cite{runge1984density, vanLeeeuven1999} and for contributions to the open question whether the required assumptions hold for systems with Coulomb interaction see   \cite{marques2012fundamentals} [Section 4.4.2] and \cite{lewin2016tddft}.} $v_{\rm xc}$ such that
the solution $\Phi(t)$ of the time-dependent Schr\"odinger equation 
\begin{align}
\begin{dcases} i \partial_t \Phi(t) = H_{\rm eff}(t) \Phi(t) \quad \mbox{for $t>0$,} \\
\Phi(0) = \Phi_0, \end{dcases}, \label{eq:nonintTDSE}
\end{align}
where the effective non-interacting Hamiltonian is given by
\begin{align}
	H_{\rm eff}(t) = -\frac12 \Delta + \sum_{j=1}^N v(r_j) + \epsilon f(t) v_{\mathcal{P}}(r_j) + \rho^{{\Phi}}(t)\ast \frac{1}{|\cdot|} (r_j) + v_{\rm xc}[\rho^{{\Phi}};\Psi_0;\Phi_0](t,r_j) , \label{eq:Heff}
\end{align}
%\textcolor{red}{Gero: I changed the red letters in the equation above from $\Psi$ to $\Phi$.}
reproduces the time-dependent electronic density $\rho^{\Psi}$ of the solution $\Psi(t)$ of the interacting evolution \eqref{eq:TD Schrodinger}. The potential $v_{\rm xc}$ at time $t$ depends not just on the density $\rho^\Psi$ at time $t$, but on its past history at all times $t'\in[0,t]$. The initial state of the non-interacting system $\Phi_0$ can be chosen arbitrarily as long as \cite[Chapter 4]{marques2012fundamentals} it reproduces the initial density $\rho^{\Psi_0}$ and the initial divergence of the current-density
\begin{align*} 
\nabla \scpr j^{\Psi_0}(r)=  N\nabla \scpr \mathrm{Im} \int_{(\R^3)^{{N-1}}} \overline{\Psi_0(r,r_2,...,r_N)} \nabla_r \Psi_0(r,r_2,...,r_N) \mathrm{d} r_1... \mathrm{d}r_N.
\end{align*}
Note however that the exact xc-potential depends on this choice.  

\subsection{Formal derivation of the Dyson equation}

In typical applications of linear response theory, the state $\Psi_0$ is the ground-state of the static interacting Hamiltonian governing evolution~\eqref{eq:TD Schrodinger}. In particular, for $\varepsilon =0$ the time-dependent density $\rho^{\Psi}$ satisfies $\rho^{\Psi}(t) = \rho^{\Psi_0}$ for all times $t\geq 0$. Consequently, if we choose the ground-state of the non-interacting system $\Phi_0$ to be the exact Kohn-Sham ground-state reproducing the density $\rho^{\Psi_0}$ (assuming it exists), the time-dependent xc-potential reduces to the exact xc-potential of static DFT, i.e. 
\begin{align*}
	v_{\rm xc}[\rho^{\Psi};\Psi_0;\Phi_0] = v_{\rm xc}^{\rm static}[\rho^{\Psi_0}],
\end{align*}
where $\Phi_0$ is the ground-state of the non-interacting Kohn-Sham Hamiltonian
\begin{align}
	H_0 = -\frac12 \Delta + \sum_{j=1}^N v(r_j) + \rho^{\Psi_0} \ast \frac{1}{|\cdot|}(r_j) + v_{\rm xc}^{\rm static}[\rho^{\Psi_0}](r_j). \label{eq:HKS}
\end{align}
From this observation we can now derive the Dyson equation of TDDFT. To this end, let us assume that the function $\rho \mapsto v_{\rm xc}[\rho ;\Phi_0;\Psi_0]$ is differentiable with respect to time-dependent densities at the stationary density $\rho(t) = \rho^{\Psi_0}$ for all $t$. Then using the expansion of the electronic density given in eq.~\eqref{eq:variation_rho} we have
\begin{align}
 v_{\rm xc}[\rho^{\Psi};\Psi_0;\Phi_0](r,t) =  v_{\rm xc}^{\rm static}[\rho^{\Psi_0}](r) + \varepsilon F_{\rm xc} (\chi \star f v_\mathcal{P}) (r,t) + \mathcal{O}(\varepsilon^2), \label{eq:xcexp}
\end{align}
where $F_{\rm xc} = \frac{\delta v_{\rm xc}}{\delta \rho} [\rho^{\Psi_0};\Psi_0;\Phi_0]$ is a linear operator whose (Schwartz) kernel is called the (exact) exchange-correlation kernel of TDDFT. Similarly, the Hartree term can be expanded in powers of $\varepsilon$ and the effective Hamiltonian \eqref{eq:Heff} becomes
\begin{align*}
H_{\rm eff}(t) = H_0 + \sum_{j=1}^N \varepsilon \bigr(F_H + F_{\mathrm{xc}}\bigr) (\chi \star f v_{\mathcal{P}})(r_j,t) + \varepsilon f(t) v_{\mathcal{P}}(r_j) + \mathcal{O}(\varepsilon^2),
\end{align*}
where $H_0$ is the Kohn-Sham Hamiltonian \eqref{eq:HKS} and $F_H$ is the Hartree operator, acting on time-dependent potentials $v$ by time-instantaneous convolution in space against the Coulomb potential,  
$$
    (F_H v)(t,r) = \frac{1}{|\cdot|}* v(t,\cdot)(r) = \int_{\R^3}\frac{1}{|r-r'|}v(t,r')\, dr'.
$$
The equivalence of the densities and \eqref{eq:variation_rho} now yields
\begin{align*}
	(\chi \star f v_{\mathcal{P}}) &= \lim_{\varepsilon \ra 0^+} \frac{\rho^{\Psi(t)} - \rho^{\Psi_0}}{\varepsilon}   = \lim_{\varepsilon \ra 0^+} \frac{\rho^{\Phi(t)} - \rho^{\Phi_0}}{\varepsilon} = \chi_0 \star f v_{\mathcal{P}} + \chi_0 \star \bigr(F_H + F_{\rm xc}\bigr) (\chi \star f v_{\mathcal{P}}),
\end{align*}
where $\chi$ and $\chi_0$ are the density-density response functions of the interacting Hamiltonian and the non-interacting Kohn-Sham Hamiltonian, respectively. As the identity above holds for every time-dependent potential $f(t) v_{\mathcal{P}}(r)$, we {obtain the celebrated Dyson equation of TDDFT}
\begin{align}
	\chi = \chi_0 + \chi_0 \star \bigr(F_H + F_{\rm xc}\bigr)  \chi . \label{eq:TDDFTDyson}
\end{align}
\begin{note} Recalling that $\chi$ is a time-dependent operator on potentials, and making the convolution in time in eq.~\eqref{eq:TDDFTDyson} and the Hartree operator explicit, the equation says that  for any time $t\geq 0$ and any potential $v_{\mathcal{P}} \in L^\infty(\R^3)$,
\begin{align*}
\chi(t) v_{\mathcal{P}} = \chi_0(t) v_{\mathcal{P}}  + \int_0^t \chi_0(t-s) \frac{1}{|\cdot|}*(\chi(s) v_{\mathcal{P}}) + \chi_0(t-s) F_{\rm xc}\bigr(\chi v_{\mathcal{P}})(s) \mathrm{d} s,
\end{align*}
where $\chi v_{\mathcal{P}}$ is the continuous $L^1(\R^3)$-valued map $s \mapsto \chi(s) v_{\mathcal{P}}$.
\end{note}

\subsection{Common approximations}
In the absence of any explicit description of the 
exact time-dependent xc potential $v_\mathrm{xc}$, all practical TDDFT calculations must resort to approximations. %Provided that the Kohn-Sham Hamiltonian \eqref{eq:HKS} satisfies Assumption~\ref{assump:groundstate}, the particular approximation chosen for the ground-state xc-potential $v_{\rm xc}^{\rm static}[\rho^{\Psi_0}]$ is immaterial to the results of this paper. 
%Regarding the approximations to the xc-kernel,} 
The two most common ones are: 
%The Runge-Gross potential is usually given by approximations of the potential in the static case where there has been an extensive research of a universal potential reproducing the exact electronic density.
%They are usually written as 

1. {\bf Random phase approximation (RPA)}: Electron-electron interaction is only taken into account on a mean field level, that is to say, in \eqref{eq:nonintTDSE}--\eqref{eq:Heff} one only keeps the Hartree term but takes $v _{xc}=0$. It follows that $F_{\rm xc} =0$ and the Dyson equation \eqref{eq:TDDFTDyson} reduces to the RPA-Dyson equation
\begin{equation} \label{A2:rpa}
   \chi  = \chi_0 + \chi_0 \star F_H \chi 
\end{equation}
treated in this paper.

2. {\bf Adiabatic local density approximation (ALDA):}
in \eqref{eq:nonintTDSE}--\eqref{eq:Heff} one uses the instantaneous (or static) local density approximation at each time $t$, 
\begin{equation} \label{ALDApot}
   v_\mathrm{xc}[\rho^\Phi;\Psi_0;\Phi_0](r,t) = v_\mathrm{xc}^{\rm LDA}[\rho^{\Phi(t)}](r)
\end{equation}
where $v_\mathrm{xc}^{\rm LDA}$ is the LDA {\it exchange-correlation potential}, that is to say 
$v_\mathrm{xc}^{\rm LDA}[\rho](r)=e'_{\rm xc}(\rho(r))$, $e_{xc}(\rhobar)$ is the exchange-correlation energy density per unit volume of a homogeneous electron gas with density $\rhobar$ (known accurately from asymptotic and numerical results \cite{CeperleyQMCcorrelation,PW92}), and $e'_{\rm xc}(\rhobar)=\frac{d}{d\rhobar}e_{\rm xc}(\rhobar)$.
Since $F_{xc}$ is the functional derivative of the exchange-correlation potential with respect to the density, it is given in this case by the multiplication operator
\begin{equation}\label{eq:ansatz_rg_potential}
    F_{\rm xc}^{\rm ALDA} (f v_{\mathcal{P}})(t,r) = \frac{\delta v_{\rm xc}^{\rm LDA}}{\delta \rho}[\rho^{\Phi_0}] (f v_{\mathcal{P}})(r,t) =  e'{}{'}_{\rm xc}\bigr(\rho^{\Phi_0}(r)\bigr) f(t) v_{\mathcal{P}}(r).
    %(_\mathrm{rg}^{\rm approx}[\{\rho(\cdot,s)\}_{s>0},\Phi_0,t] = \rho(\cdot,t) * \frac{1}{|\cdot|} + v_\mathrm{xc}[\rho(\cdot,t)]  + v(r),
\end{equation}

%\textcolor{red}{Gero: @Thiago, I like your idea to write out the xc operator explicitly, so I kept it, but you had one derivative missing in the rightmost expression.}
%Refined static exchange-correlation potentials, e.g. of the form $v_\mathrm{xc}^{\rm GGA}(r)=f(\rho(r),\nabla\rho(r))$, could also be considered. 
All the results in this paper (Theorems~\ref{thm:time_rpa_solution},~\ref{theo:polesRPA} and~\ref{theo:rankpolesRPA}) apply to eq.~\eqref{A2:rpa}. Note that the particular approximation chosen for the static xc potential $v_{\rm xc}^{\rm static}$ underlying the reference response function $\chi_0$ can be arbitrary, provided the Hamiltonian 
\eqref{eq:HKS} satisfies Assumption~\ref{assump:groundstate}. 

In the approximation~\eqref{ALDApot}, the time-dependent xc-potential only depends instantaneously on the electronic density $\rho(\cdot,t)$. This is an  adiabatic approximation and it is one of the main challenging problems in TDDFT to improve on it in a systematic way \cite[Chap. 8]{ullrich2012time}.

\section{Poles of the density-density response function of a non-interacting Hamiltonian}
\label{appendix:poles_noninteractingDDRF}

\begin{prop}
    \label{prop:chi_zero} 
    Let $\chi$ be the density-density response function defined in \eqref{eq:responseoperatordef} of a non-interacting Hamiltonian $H$ satisfying Assumption~\ref{assump:groundstate}, \emph{i.e.} $H = -\frac{1}{2}\Delta + \sum_{i=1}^N v(r_i)$. 
    Let $h$ be the one-body Hamiltonian $h=-\frac{1}{2}\Delta + v$ and $(\epsilon_k,\psi_k)_{1 \leq k \leq N} \in \big(\mathbb{R}\times L^2(\mathbb{R}^3))^N$ be its lowest eigenpairs. 
    Then we have for all $v_\mathcal{P} \in L^2(\mathbb{R}^3)+L^\infty(\mathbb{R}^3)$ and $z \in \mathbb{C}$ with $\mathrm{Im}(z)>0$
    \begin{equation}
      \label{eq:chi0}
      \widehat{\chi}(z) v_\mathcal{P}(r) = \sum\limits_{k=1}^N \overline{\psi_k}(r) \big[ (\epsilon_k-z-h)^{-1} + (\epsilon_k+z-h)^{-1}\big](v_\mathcal{P}\psi_k)(r).
    \end{equation}
  \end{prop}
  
  \begin{proof}
    \textbf{Step 1.} We first show that 
    \begin{multline}
      \label{eq:awful1}
      \big((E_0-z-H)^{-1} + (E_0+z-H)^{-1}\big) S^* v_\mathcal{P} \\
      = \frac{1}{\sqrt{N!}} \sum\limits_{j=1}^N \sum\limits_{\sigma \in S_N} (-1)^\sigma \left[(\epsilon_{\sigma(j)}-z-h)^{-1} + (\epsilon_{\sigma(j)}+z-h)^{-1}\right] v_\mathcal{P} (r_j) \prod_{k=1}^N\psi_{\sigma(k)}(r_k).
    \end{multline}
  
    Since $H\Psi_0 = E_0 \Psi_0$, we have
    \begin{equation}
      \big((E_0-z-H)^{-1} + (E_0+z-H)^{-1}\big) S^* v_\mathcal{P} = \big((E_0-z-H)^{-1} + (E_0+z-H)^{-1}\big) \sum\limits_{i=1}^N v_\mathcal{P}(r_i) \Psi_0(r_1,\dots,r_N).
    \end{equation}
    Using that for any bounded continuous $g$, we have 
    \begin{equation}
        \int_\mathbb{R} g(\lambda) \, \mathrm{d}P^{H}_\lambda = \int_{\mathbb{R}^N} g(\mu_1 + \dots + \mu_N) \, \mathrm{d}P^h_{\mu_1} \otimes \cdots \otimes \mathrm{d}P^h_{\mu_N},
      \end{equation}
      and noticing that for $\sigma \in S_N$, $E_0 = \sum\limits_{k=1}^N \epsilon_{\sigma(k)}$ we get \eqref{eq:awful1}.
  
    \textbf{Step 2.} We have
    \begin{align}
      S& \big((E_0-z-H)^{-1} + (E_0+z-H)^{-1}\big) S^* v_\mathcal{P}(r) \\
      &= \frac{N}{\sqrt{N!}} \int_{\mathbb{R}^{3(N-1)}} \overline{\Psi_0}(r,r_2,\dots,r_N)  \nonumber \\
      & \qquad \sum\limits_{\sigma \in S_N} (-1)^\sigma \left[(\epsilon_{\sigma(j)}-z-h)^{-1} + (\epsilon_{\sigma(j)}+z-h)^{-1}\right] \big(v_\mathcal{P}(r) \psi_{\sigma(1)}(r)\big)\prod_{k \geq 2}\psi_{\sigma(k)}(r_k) \, \mathrm{d}r_2 \dots \, \mathrm{d}r_N  \nonumber\\ 
      & \qquad - \Big\langle \Psi_0, \frac{1}{\sqrt{N!}} \sum\limits_{j=1}^N \sum\limits_{\sigma \in S_N} (-1)^\sigma \left[(\epsilon_{\sigma(j)}-z-h)^{-1} + (\epsilon_{\sigma(j)}+z-h)^{-1}\right] v_\mathcal{P}(r_j) \prod_{k=1}^N\psi_{\sigma(k)}(r_k) \Big\rangle \rho_0(r)
    \end{align}
    %\textcolor{red}{Gero: I think it should be $\overline{\Psi_0}$ above}
    The second term on the RHS of the above equation vanishes since $(h-\epsilon_{\sigma(j)})\psi_{\sigma(j)} = 0$. 
    Thus by orthonormality of $(\psi_i)_{1 \leq i \leq N}$, we obtain:
    \begin{equation}
        \widehat{\chi}(z) v_\mathcal{P}(r) = \sum\limits_{k=1}^N \overline{\psi_k}(r) \big[ (\epsilon_k-z-h)^{-1} + (\epsilon_k+z-h)^{-1}\big](v_\mathcal{P}\psi_k)(r).
    \end{equation}
  \end{proof}
  %\textcolor{red}{Gero: I think the first of the two $\psi_k$'s above is a $\overline{\psi_k}$.}
  
    We notice that the expression~\eqref{eq:chi0} is equivalent to the {Lehmann representation} typically found in the physics and chemistry literature if, as is often assumed in this literature (but not strictly speaking valid for molecular Hamiltonians due to the presence of continuous spectrum), $h$ is diagonalizable in an orthonormal basis $(\psi_i)_{i \in \mathbb{N}}$.  Under this assumption,
    \begin{multline*}
      \big[ (\epsilon_k-z-h)^{-1} + (\epsilon_k+z-h)^{-1}\big](v_\mathcal{P}\psi_k)(r) \\ 
      = \sum\limits_{a=1}^\infty \Big( \frac{1}{\epsilon_k-\epsilon_a-z} + \frac{1}{\epsilon_k-\epsilon_a+z} \Big)  \langle \psi_a ,v_\mathcal{P} \psi_k\rangle  \psi_a(r).
    \end{multline*}
    %\textcolor{red}{Gero: I think the matrix element above should be $\langle \psi_a,v_{\mathcal P}\psi_k\rangle$.}  
    Hence for $v_\mathcal{O} \in L^2(\mathbb{R})+L^\infty(\mathbb{R}^3)$, we obtain
    \begin{eqnarray}
      \langle v_\mathcal{O}, \widehat{\chi}(z) v_\mathcal{P} \rangle &=& \sum\limits_{k=1}^N \sum\limits_{a=1}^\infty \Big( \frac{1}{\epsilon_k-\epsilon_a-z} + \frac{1}{\epsilon_k-\epsilon_a+z} \Big) \langle \psi_k v_\mathcal{O}, \psi_a  \rangle \langle \psi_a ,v_\mathcal{P}\psi_k \rangle  \nonumber \\ 
      &=& \sum\limits_{k=1}^N \sum\limits_{a=N+1}^\infty \Big( \frac{1}{\epsilon_k-\epsilon_a-z} + \frac{1}{\epsilon_k-\epsilon_a+z} \Big) \langle \psi_k v_\mathcal{O}, \psi_a \rangle \langle \psi_a,v_\mathcal{P} \psi_k \rangle. 
      \label{lehm_nonint}
    \end{eqnarray}
    %\textcolor{red}{Gero: I think the matrix elements above should be $\langle \psi_k,v_{\mathcal O}\psi_a\rangle \langle \psi_a,v_{\mathcal P}\psi_k\rangle$.}
    {Note that the Lehmann representation \eqref{lehm_nonint} could also have been easily obtained from \eqref{lehmann1}, by using the fact that the excited states of the non-interacting Hamiltonian $H$ are given by the Slater determinants of the eigenstates of $h$ containing at least one unoccupied eigenstate $\psi_a$, $a>N$, and only those Slater determinants containing exactly one unoccupied eigenstate yield nonzero matrix elements in \eqref{lehmann1}}. 
    
    Eq.~\eqref{lehm_nonint} shows that the poles of the density-density response function of a non-interacting Hamiltonian are located at the spectral gaps between occupied and virtual eigenvalues of the one-body Hamiltonian.

\section{Spectral theory of bounded operators}

We collect here some results on the spectral theory of bounded operators on Banach spaces. These results are elementary and complete proofs can be found in \cite{gohberg2013classes}. For convenience of the reader, we briefly sketch these proofs here.

\begin{lem}[Continuity of Spectra]\label{lem:continuityofspectra} Let $A \in \mathcal{B}(E)$ where $E$ is a Banach space and $\mu \in \rho(A)$ (where {$\mathcal{B}(E)$ denotes the set of bounded linear operators on $E$ and} $\rho(A)$ denotes the resolvent set of $A$). Then for any $B$ with $\norm{B} < \norm{(\mu-A)^{-1}}$ we have $\mu \in \rho(A+B)$. In particular, if $A: B_\delta(0)\subset \C \rightarrow \mathcal{B}(E)$ is continuous and $W \subset \rho\bigr(A(0)\bigr)$ is compact, then $W \subset \rho\bigr(A(z)\bigr)$ for any $z$ close enough to $0$.
\end{lem}

\begin{proof} For $\mu \in \rho(A)$, we have
\begin{align*}
    (\mu - A)^{-1} (\mu - A - B) = I - (\mu - A)^{-1} B \quad \mbox{and}\quad (\mu - A - B) (\mu - A)^{-1} = I - B (\mu - A)^{-1}
\end{align*}
Thus, for $\norm{B} < \norm{(\mu-A)^{-1}}^{-1}$ the operators above are of the form $I-K$ with $\norm{K}<1$. Hence, they are invertible and the inverse is given by the Neumann series, $\sum_{k\in \N} K^n$.

For the second statement, note that since $W$ is compact, and $(\mu - A(0))^{-1}$ is continuous with respect to $\mu$, we have $\epsilon = \inf_{\mu \in W} \norm{\bigr(\mu - A(0)\bigr)^{-1}}^{-1} > 0$. Thus, from continuity we have $\norm{A(z) - A(0)} \leq \epsilon$ for any $z$ close to $0$ and the result follows from the first statement.
\end{proof}

\begin{lem}[Riesz projection and separation of spectra] Let $\gamma \subset \rho(A) \subset \C$ be a closed smooth curve separating the spectrum of $A$. Then, the operator
\begin{align*}
    P = \frac{1}{2\pi i} \oint_{\gamma}(\mu- A)^{-1} d\mu
\end{align*}
is a projection commuting with $A$. Moreover, for $\mu_0 \not \in  \gamma$, the operator 
\begin{align}
    S(\mu_0) =\frac{1}{2\pi i} \oint_{\gamma} \frac{1}{\mu- \mu_0}(\mu - A)^{-1}(1-P) \mathrm{d} \mu . \label{eq:inverse1-Pdef}
\end{align}
satisfies
\begin{align}
    S(\mu_0) = \begin{dcases} \bigr((1-P)(\mu_0 - A)(1-P)\bigr)^{-1}, &\mbox{for $\mu_0$ inside $\gamma$,}\\
    \bigr(P(\mu_0 - A) P\bigr)^{-1}, &\mbox{for $\mu_0$ outside $\gamma$.} \end{dcases} \label{eq:inverse1-Pformula}
\end{align}
In particular, the spectrum of $A\bigr\rvert_{\ran P} \in \mathcal{B}(\ran P)$ and $A \bigr \rvert_{\ker P} \in \mathcal{B}(\ker P)$ is given respectively by the spectrum of $A$ inside and outside of $\gamma$.
\end{lem}

\begin{proof} That the operator $P$ is well-defined and bounded is clear since $\gamma \subset \rho(A)$ and $\mu \mapsto (\mu - A)^{-1}$ is continuous in $\mu$. To see that $P$ is a projection, note that one can choose a curve $\gamma_1$ inside $\gamma$ such that all points lying between $\gamma_1$ and $\gamma$ are in the resolvent of $A$. Thus from a standard argument of holomorphic function theory, 
\begin{align*}
    P = \frac{1}{2\pi i} \oint_{\gamma_1} (\lambda - A)^{-1} \mathrm{d} \lambda.
\end{align*}
Hence, multiplying the above by the definition of $P$ with a contour integral on $\gamma$, using the resolvent identity $(\mu - A)^{-1}(\lambda - A)^{-1} = (\mu-\lambda)^{-1}\bigr((\lambda - A)^{-1}- (\mu-A)^{-1}$ and using the Cauchy integral formula for holomorphic functions, one can show that $P^2 = P$. 

Since $S(\mu_0)$ commutes with $A$, formula \eqref{eq:inverse1-Pformula} follows from
\begin{align*}
    (\mu_0 - A) S(\mu_0) &= \frac{1}{2\pi i} \oint_{\gamma} \frac{1}{\mu - \mu_0} - \frac{1}{2\pi i} \oint_{\gamma} (\mu - A)^{-1} \mathrm{d} \mu = \begin{dcases} 1-P, &\mbox{for $\mu_0$ inside $\gamma$, } \\
    -P, &\mbox{for $\mu_0$ outside $\gamma$,} \end{dcases}
\end{align*}
and 
\begin{align*}
    S(\mu_0) P &= \frac{1}{2\pi i} \oint_{\gamma} \oint_{\gamma_1} \frac{(\mu-A)^{-1} - (\lambda -A)^{-1}}{(\mu-\mu_0)(\lambda - \mu)} \mathrm{d}\lambda \mathrm{d}\mu = \begin{dcases} 0, &\mbox{for $\mu_0$ inside $\gamma$, }\\
    S(\mu_0), &\mbox{for $\mu_0$ outside $\gamma$.}\end{dcases}
\end{align*}
    
Finally, the last statement follows from two observations. First, the existence of the inverses in \eqref{eq:inverse1-Pformula} implies that $\sigma( A \bigr \rvert_{\ran P})$ lies inside $\gamma$ and $\sigma(A\bigr\rvert_{\ker P})$ lies outside $\gamma$. Second, from the decomposition 
\begin{align*}
    A = \begin{pmatrix} A\bigr\rvert_{\ran P} & 0 \\
    0 & A\bigr\rvert_{\ker P} \end{pmatrix}
\end{align*}
with respect to $\mathcal{H} = \ran P \oplus \ker P$, we have $\sigma(A) = \sigma(A\bigr\rvert_{\ker P}) \cup \sigma(A\bigr \rvert_{\ran P})$.
\end{proof}

\bibliographystyle{alpha}
\bibliography{linear-response}
\end{document}